\documentclass[12pt]{article}
\usepackage{amssymb,amsmath,amsfonts,latexsym}
\usepackage{natbib}
\usepackage{color}
\usepackage{colortbl,multirow}
\usepackage{graphicx}
\usepackage{epstopdf}
\AppendGraphicsExtensions{.tif}

\def\R{\ensuremath{\mathbb R}}

\def\Y{\mathcal{Y}}

\def\1{{\bf 1}}

\def\o{\ensuremath{\underline{\omega}}}

\begin{document}
\title{Predicting Climate Change using Response Theory: Global Averages and Spatial Patterns\footnote{Paper prepared for the special issue of the Journal of Statistical Physics dedicated to the $80^{th}$ birthday of Y. Sinai and D. Ruelle.}}
\author{ Valerio Lucarini$^{1,2}$ [\texttt{valerio.lucarini@uni-hamburg.de}]\\
$^1$ Meteorologisches Institut, CEN, University of Hamburg\\
Hamburg, Germany\\
$^2$ Department of Mathematics and Statistics\\
University of Reading, Reading, UK \\
\\
Francesco Ragone$^3$\\
$^3$ Laboratoire de Physique de l'\'{E}Ecole Normale \\Sup\'{e}erieure de Lyon
Lyon, France\\
\\
Frank Lunkeit$^1$\\
$^1$ Meteorologisches Institut, CEN, University of Hamburg\\
Hamburg, Germany}

\date{\today}
\maketitle
\abstract{The provision of accurate methods for predicting the climate response to anthropogenic and natural forcings is a key contemporary scientific challenge. Using a simplified and efficient open-source general circulation model of the atmosphere featuring O($10^5$) degrees of freedom, we show how it is possible to approach such a problem using nonequilibrium statistical mechanics. Response theory allows one to practically compute the time-dependent measure supported on the pullback attractor of the climate system, whose dynamics is non-autonomous as a result of time-dependent forcings. We propose a simple yet efficient method for predicting  - at any lead time and in an ensemble sense - the change in climate properties resulting from increase in the concentration of CO$_2$ using test perturbation model runs. We assess strengths and limitations of the response theory in predicting the changes in the globally averaged values of surface temperature and of the yearly total precipitation, as well as in their spatial patterns. The quality of the predictions obtained for the surface temperature fields is rather good, while in the case of precipitation a good skill is observed only for the global average. We also show how it is possible to define accurately concepts like the the inertia of the climate system or to predict when climate change is detectable given a scenario of forcing. Our analysis can be extended for dealing with more complex portfolios of forcings and can be adapted to treat, in principle, any climate observable. Our conclusion is that climate change is indeed a problem that can be effectively seen through a statistical mechanical lens, and that there is great potential for optimizing the current coordinated modelling exercises run for the preparation of the subsequent reports of the Intergovernmental Panel for Climate Change.}
\section{Introduction}
%\subsection{A Brief Summary of Response Theory}
The climate is a forced and dissipative nonequilibrium chaotic system with a complex natural variability resulting from the  interplay of instabilities and re-equilibrating mechanisms, negative and positive feedbacks, all covering a very large range of spatial and temporal scales.  One of the outstanding scientific challenges of the last decades has been the attempt to provide a comprehensive theory of climate, able to explain the main features of its dynamics, describe its variability, and predict its response to a variety of forcings, both anthropogenic and natural \cite{ghil_topics_1987,LBHRPW14,Ghil2015}. The study of the phenomenology of the climate system is commonly approached by focusing on distinct aspects like:
\begin{itemize}
\item wave-like features such Rossby  or equatorial waves, which have enormous importance in terms of predictability and transport of, \textit{e.g.}, energy, momentum, and water vapour;
\item particle-like features such as hurricanes, extratropical cyclones, or oceanic vortices, which are of great relevance for the local properties of the climate system and its subdomains; 
\item turbulent cascades, which determine, \textit{e.g.} dissipation in the boundary layer and development of large eddies through the mechanism of geostrophic turbulence. 
\end{itemize} 
While each of these points of view is useful and necessary, none is able to provide alone a comprehensive understanding of the properties of the climate system{; see also discussion in \cite{Ghil02}}.

On a macroscopic level, one can say that at zero order the climate is driven by differences in the absorption of solar radiation across its domain. The prevalence of absorption at surface and at the lower levels of the atmosphere leads, through a rich portfolio of processes, to compensating vertical energy fluxes (most notably, convective motions in the atmosphere and exchanges of infrared radiation), while the prevalence of absorption of solar radiation in the low latitudes regions leads to the set up of the large scale circulation of the atmosphere (with the hydrological cycle playing a key role), which allows for reducing the temperature differences between tropics and polar regions with respect to what would be the case in absence of horizontal energy transfers \cite{Lor67,Peixoto:1992}. 

It is important to stress that such organized motions of the geophysical fluids, which act as negative feedbacks but cannot be treated as diffusive Onsager-like processes, typically result from the transformation of some sort of available potential into kinetic energy, which contrasts the damping due to a variety of dissipative processes. { See \cite{Lucarini2014a} for a detailed analysis of the relationship between response, fluctuations, and dissipation at different scales.}  Altogether, the climate can be seen as a thermal engine able to transform heat into mechanical energy with a given efficiency, and featuring many different irreversible processes that make it non-ideal \cite{Kleidon05,Lucarini09PRE,LBHRPW14}. 

Besides the strictly scientific aspect, much of the interest on climate research has been driven in the past decades by the accumulated observational evidence of the human influence on the climate system. In order to summarize and coordinate the research activities carried on by the scientific community, the United Nations Environment Programme (UNEP) and the World Meteorological Organization (WMO) established in 1988 the International Panel on Climate Change program (IPCC). The IPCC reports, issued periodically about every 4-5 years, provide systematic reviews of the scientific literature on the topic of climate dynamics, with special focus on global warming and on the socio-economic impacts of anthropogenic climate change \cite{IPCC01,IPCC07,IPCC13}. Along with such a review effort, the IPCC defines standards for the modellistic exercises to be performed by research groups in order to provide projections of future climate change with numerical models of the climate system. A typical IPCC-like climate change experiment consists in simulating the system in a reference state (a stationary preindustrial state with fixed parameters, or a realistic simulation of the present-day climate), raising the $CO_2$ concentration (as well as, in general, the concentration of other greenhouse gases such as methane) in the atmosphere following a certain time modulation in a certain time window, and then fixing the $CO_2$ concentration to a certain value to observe the relaxation of the system to a new stationary state. Each time-modulation of the $CO_2$ forcing defines a \textit{scenario}, and it is a representation of the expected $CO_2$ increase resulting from a specific path of industrialization and change in land use. { Note that the attribution of unusual climatic conditions to  specific climate forcing is far from being a trivial matter  \cite{Allen2003,Hannart2016}.}

While much progress has been achieved, we are still far from having a conclusive framework of climate dynamics. One needs to consider that the study of climate faces, on top of all the difficulties that are intrinsic to any nonequilibrium system, the following additional aspects that make it especially hard to advance its understanding:
\begin{itemize}
\item the presence of well-defined subdomains - the atmosphere, the ocean, \textit{etc.} - featuring extremely different physical and chemical properties, dominating dynamical processes, and characteristic time-scales;
\item the complex processes coupling such subdomains;
\item the presence of a continuously varying set of forcings resulting from, \textit{e.g.}, the fluctuations in the incoming solar radiation and the processes - natural and anthropogenic - altering the atmospheric composition;
\item the lack of scale separation between different processes, which requires a profound revision of the standard methods for model reduction/projection to the slow manifold, and calls for the unavoidable need of complex parametrization of subgrid scale processes in numerical models;  
\item the impossibility to have detailed and homogeneous observations of the climatic fields with extremely high-resolution in time and in space, and the need to integrate direct and indirect measurements when trying to reconstruct the past climate state beyond the industrial era; 
\item the fact that we can observe only one realization of the process. 
\end{itemize}  

Since the climate is a nonequilibrium system, it is far from trivial to derive its response to forcings from the natural variability realized when no time-dependent forcings are applied. { This is the fundalmental reason why the construction of a theory of the climate response is so challenging \cite{Ghil2015}}. As already noted by Lorenz \cite{lorenz79}, it is hard to construct a one-to-one correspondence between forced and free fluctuations in a climatic context. Following the pioneering contribution by Leith \cite{Leith75}, previous attempts on predicting climate response based broadly on applications of the fluctuation-dissipation theorem have had some degree of success \cite{langen_estimating_2005,gritsun_climate_2007,majda07,gritsun2008b,cooper_climate_2011}, but, in the deterministic case, the presence of correction terms due to the singular nature of the  invariant measure make such an approach potentially prone to errors \cite{R09,LS11}. Adding noise in the equations in the form of stochastic forcing - as in the case of using stochastic parametrizations \cite{Fetal15} in  multiscale system - provides a way to regularize the problem, but it is not entirely clear how properties convergence in the zero noise. Additionally, one should provide a robust and meaningful construction of the model to be used for constructing the noise: a proposal in this direction is given in  \cite{WL12,WL13,LBHRPW14}.

In this paper we want to show how climate change is indeed a well-posed  problem at mathematical and physical level by presenting a theoretical analysis of how PLASIM \cite{plasim}, a  general circulation model of intermediate complexity responds to simplified yet representative changes in the atmospheric composition mimicking increasing concentrations of greenhouse gases. We will frame the problem of studying the statistical properties of a non-autonomous, forced and dissipative complex system using the mathematical construction of the { pullback attractor \cite{GCS08,CSG11,CLR13} - see also  the closely related concept of snapshot attractor \cite{BKT11,BT12,BKT13,DBT15} -} and will use as theoretical framework the Ruelle response theory \cite{R98,R09} to compute the effect of small time-dependent perturbations on the background state. We will stick to the linear approximation, which has proved its effectiveness in various examples of geophysical interest \cite{LS11,RLL14}. The basic idea is to use a set of probe perturbations to derive the Green function of the system, and be then able to predict the response of the system to a large class of forcings having the same spatial structure. In this way, the uncertainties associated to the application of the fluctuation-dissipation theorem are absent and the theoretical framework is more robust. Note that, as shown in \cite{L09}, one can practically implement the response theory also to treat the nonlinear effects of the forcing. 

PLASIM  has O$(10^5)$ degrees of freedom and provides a flexible tool for performing theoretical studies in climate dynamics. PLASIM is much faster (and indeed simpler) than the state of the art climate models used in the IPCC reports, but provides a reasonably realistic representation of atmospheric dynamics and of its interactions with the land surface and with the mixed layer of the ocean. The model includes a suite of efficient parametrization of small scale processes such as those relevant for describing radiative transfer, clouds formation, and turbulent transport across the boundary layer. { It is important to recall that in climate science it is practically necessary and conceptually very sound to choose models belonging to different levels of a hyeararchy of complexity \cite{DG05} depending on the specific problem to be studied: the most physically comprehensive and computationally expensive model is \textit{not} the best model to be used for all purposes; see discussion in \cite{lucarini_modelling_2013}.} 

In a previous work \cite{RLL14} we have considered a somewhat unrealistic set-up for the model, where the meridional  oceanic heat transport was set to zero, with no feedback from the climate state. Such a limitation resulted in an extremely high  increase of the globally averaged surface temperature resulting from higher $CO_2$ concentrations. In this paper we extend the previous analysis by using a better model and by considering a wider range of climate observables able to provide a more complete picture of the climate response to $CO_2$ concentration. In particular, we wish to show to what extent response theory is suited for performing projections of the spatial pattern of climate change. 

The paper is organized as follows. In Section \ref{pullback} we introduce the main concepts behind the theoretical framework of our analysis. We  briefly describe the basic properties of the pullback attractor and explain its relevance in the context of climate dynamics. We then discuss the relevance of response theory for studying situations where  the non-autonomous dynamics can be decomposed into a dominating autonomous component plus a small non-autonomous correction. In Section \ref{methods} we introduce the climate model used in this study, discuss the various numerical experiments and the climatic observables of our interest, and present the data processing methods used for predicting the climate response to forcings. In Section \ref{results} we present the main results of our work. We focus on  two observables of great relevance, namely the surface temperature and the yearly total precipitation, and investigate the skill of response theory in predicting the change in their statistical properties, exploring both changes in global quantities and spatial patterns of changes. We will also show how to flexibly use response theory for predicting when climate change becomes statistically significant in a variety of scenarios.  In Section \ref{conclusions} we summarize and discuss the main findings of this work. In Section \ref{future} we propose some ideas for potentially exciting future investigations.

\section{Pullback Attractor and Climate Response}\label{pullback} 
Since the climate system experiences  forcings whose variations take place on many different time scales \cite{saltzman_dynamical}, defining rigorously what climate response actually is requires some care. It seems relevant to take first a step in the direction of  considering the rather natural situation where we want to estimate the statistical properties of complex non-autonomous dynamical systems. %Time-dependence can be related to the presence of natural periodic phenomena, such as in the case of the seasonal cycle when looking at hot or cold extremes of temperatures (or of energy consumption for heating/cooling buildings), or of slow modulations to the parameters of the system, as in the eponymous case of climate change. We will present two different yet related approaches for studying this problem.

Let us then consider a continuous-time dynamical system 
\begin{equation}
\dot{x}=F(x,t)\label{eq1}
\end{equation}
on a compact manifold $\Y\subset\mathbb{R}^d$,  where $x(t)=\phi(t,t_0)x(t_0)$, with $x(t=t_0)=x_{in}\in \Y$ initial condition and $\phi(t,t_0)$ is defined for all $t\geq t_0$ with $\phi(s,s)=\1$. The two-time evolution operator $\phi$ generates a two-parameter semi-group. In the autonomous case, the evolution operator generates a one-parameter semigroup, because of time translational invariance, so that $\phi(t,s)=\phi(t-s)$ $\forall t\geq s$. In the non-autonomous case, in other terms, there is an \textit{absolute clock}. We want to consider forced and dissipative systems such that with probability one initial conditions in the infinite past are attracted at time $t$ towards $\mathcal{A}(t)$, a time-dependent family of geometrical sets. In more formal terms, we say a family of objects $\cup_{t\in\R} \mathcal{A}(t)$ in the finite-dimensional, complete metric phase space $\Y$ is a pullback attractor for the system  $\dot{x}=F(x,t)$ if the following  conditions are obeyed:
\begin{itemize}
\item $\forall t$, $\mathcal{A}(t)$ is a compact subset of $\Y$ which is covariant with the dynamics, \textit{i.e.} $\phi(s,t)\mathcal{A}(t)=\mathcal{A}(s)$, $s\geq t$.
\item $\forall t$ $\lim_{t_0\rightarrow-\infty}d_{\Y}(\phi(t,t_0)B,\mathcal{A}(t))=0$ for $a.e.$ measurable set $B\subset \Y$.
\end{itemize}
where $d_\Y(P,Q)$ is the Hausdorff semi-distance between the  $P \subset \Y$ and $Q\subset \Y$. We have that $d_\Y(P,Q)=\sup_{x\in P}d_\Y(x,Q)$, with $d_\Y(x,Q)=\inf_{y\in Q}d_\Y(x,y)$. We have that, in general, $d_\Y(P,Q)\neq d_\Y(Q,P)$ and $d_\Y(P,Q)=0\Rightarrow P\subset Q$. See a detailed discussion of these concepts in, \textit{e.g.}, \cite{GCS08,CSG11,CLR13}. Note that a substantially similar construction, the \textit{snapshot attractor}, has been proposed and fruitfully used to address a variety of time-dependent problems, including some of climatic relevance \cite{BKT11,BT12,BKT13,DBT15}.

In some cases, the geometrical set $\mathcal{A}(t)$ supports useful  measures $\mu_t(\textrm{d}x)$.  These can be obtained as evolution at time $t$ through the Ruelle-Perron-Frobenius operator \cite{R78} of the Lebesgue measure supported on $B$ in the infinite past, as from the conditions above.  Proposing a minor generalization of the \textit{chaotic hypothesis} \cite{GC95}, we assume that when considering sufficiently high-dimensional, chaotic and non-autonomous dissipative systems, at all practical levels - \textit{i.e.} when one considers macroscopic observables - the corresponding measure $\mu_t(\textrm{d}x)$ constructed as above is of the SRB type. This amounts to the fact that we can construct at all times $t$ a meaningful (time-dependent) physics for the system. Obviously, in the autonomous case, and under suitable conditions - \textit{e.g.} in the case of of Axiom A system or taking the point of view of the chaotic hypothesis -  $\mathcal{A}(t)=\Omega$ is the attractor of the system (where the $t-$ dependence is dropped), which supports the SRB invariant measure $\mu(\textrm{d}x)$.

Note that when we analyze the statistical properties of a numerical model describing a non-autonomous forced and dissipative system, we often follow - sometimes inadvertently - a protocol that mirrors precisely the definitions given above. We start many simulations in the distant past with initial conditions chosen according to an a-priori distribution. After a sufficiently long  time, related to the slowest time scale of the system, at each instant the statistical properties of the ensemble of simulations do not depend anymore on the choice of the initial conditions.  A prominent example of this procedure is given by how simulations of past and historical climate conditions are performed in the modeling exercises such as those demanded by the IPCC \cite{IPCC01,IPCC07,IPCC13}, where time-dependent  climate forcings due to changes in greenhouse gases, volcanic eruptions, changes in the solar irradiance, and other astronomical effects are taken into account for defining the radiative forcing to the system. Note that future climate projections are \textit{always} performed using as initial conditions the final states of simulations of historical climate conditions, with the result that the covariance properties of the $\mathcal{A}(t)$ set are maintained. 

Computing the expectation value of measurable observables for the time dependent measure $\mu_t(\textrm{d}x)$ resulting from the evolution of the dynamical system given in Eq. \ref{eq1} is in general far from trivial and requires constructing a very large ensemble of initial conditions in the Lebesgue measurable set $B$ mentioned before. Moreover, from the theory of pullback attractors we have no real way to predict the sensitivity of the system to small changes in the dynamics. 

The response theory introduced by Ruelle \cite{R98,R09} (see also extensions and a different point of view summarized in, \textit{e.g.} \cite{B08}) allows for computing the change in the measure of an Axiom A system resulting from weak perturbations of intensity $\epsilon$ applied to the dynamics in terms of the properties of the unperturbed system. The basic concept behind the Ruelle response theory is that the invariant measure of the system, despite being supported on a strange geometrical set, is differentiable with respect to $\epsilon$. See \cite{L15} for a discussion on the radius of convergence (in terms of $\epsilon$) of the response theory. 

In this case, instead, our focus is on saying that the Ruelle response theory allows for constructing the time-dependent measure of the pullback attractor $\mu_t(\mathrm{d}x)$ by computing the time-dependent corrections of the measure with respect to a reference state. In particular, let us assume that we can write 
\begin{equation}
\dot{x}=F(x,t)=F(x)+\epsilon X(x,t)\label{eqpertu}
\end{equation}
where $|\epsilon X(x,t)|\ll |F(x)|$ $\forall t\in \R$ and $\forall x\in \Y$, so that we can treat $F(x)$ as the background dynamics and $\epsilon X(x,t)$ as a perturbation. Under appropriate mild regularity conditions, it is  possible to perform a Schauder decomposition \cite{LT96} of the forcing, so that we express $X(x,t)=\sum_{k=1}^\infty X_k(x)T_k(t)$. Therefore, we restrict our analysis without loss of generality to the case where $F(x,t)=F(x)+\epsilon X(x)T(t)$.

One can evaluate the expectation value of a measurable observable $\Psi(x)$ on the time dependent measure $\mu_t(\textrm{d}x)$ of the system given in Eq. \ref{eq1}  as follows: 
\begin{equation}
\int \mu_t(\mathrm{d}x) \Psi(x)=\langle \Psi \rangle^\epsilon(t)=\langle \Psi \rangle_0+\sum_{j=1}^\infty \epsilon^j \langle \Psi\rangle_0^{(j)}(t)\label{mut},
\end{equation}
where $\langle \Psi \rangle_0 = \int \bar{\mu}(\mathrm{d}x) \Psi(x)$ is the expectation value of $\Psi$ on the SRB invariant measure $\bar{\mu}(\mathrm{d}{x})$ of the dynamical system $\dot{x}=F(x)$. Each term $\langle \Psi \rangle_0^{(j)}(t)$ can be expressed as time-convolution of  the $j^{th}$ order Green function $G^{(j)}_\Psi$ with the time modulation $T(t)$:
\begin{equation}
\langle \Psi\rangle_0^{(j)}(t) =\int_{-\infty}^\infty  d\tau_1\ldots \int_{-\infty}^\infty d\tau_n G^{(j)}_\Psi(\tau_1,\ldots,\tau_j)T(t-\tau_1)\ldots T(t-\tau_j)\label{deltan2}.
\end{equation}
At each order, the Green function can be written as:
\begin{equation}
G^{(j)}_\Psi(\tau_1,\ldots,\tau_j) = \int \bar{\mu}(\mathrm{d}{x})\Theta(\tau_{j}-\tau_{j-1})\ldots \Theta(\tau_1)\Lambda S_0^{\tau_1} \ldots S_0^{\tau_{j-1}} \Lambda S_0^{\tau_{j}}  \Psi(x),
\end{equation}
where  $\Lambda(\bullet)={X} \cdot \nabla (\bullet)$ and $S_0^t(\bullet)=\exp(t {F}\cdot \nabla )(\bullet)$ is the unperturbed evolution operator while the Heaviside $\Theta$ terms enforce causality. In particular, the linear correction term can be written as:
\begin{equation}
\langle \Psi\rangle_0^{(1)}(t), = \int \bar{\mu}(\mathrm{d}{x}) \int_0^\infty \mathrm{d} \tau \Lambda S_0^{\tau}  \Psi(x) T(t-\tau)= \int_{-\infty}^\infty \mathrm{d} \tau G^{(1)}_\Psi(\tau) T(t-\tau),\label{lrruelle}
\end{equation}
while, considering the Fourier transform of Eq. \ref{lrruelle}, we have:
\begin{equation}
\langle \Psi\rangle_0^{(1)}(\omega) = \chi^{(1)}_\Psi(\omega) T(\omega),\label{lrruelle2}
\end{equation}
where we have introduced the susceptibility $\chi^{(1)}_\Psi(\omega)=\mathcal{F}[G^{(1)}_\Psi]$, defined as the Fourier transform of the Green function $G^{(1)}_\Psi(t)$. Under suitable integrability conditions, the fact that the Green function $G^{(t)}$ is causal is equivalent to saying that its susceptibility   obeys the so-called Kramers-Kronig relations \cite{ruelle2009,LS11}, which provide integral constraints linking its real and imaginary part, so that $\chi^{(1)}(\omega)=\mathrm{i} \mathcal{P}(1/\omega)\star\chi^{(1)}(\omega)$, where $i=\sqrt{-1}$, $\star$ indicates the convolution product, and $\mathcal{P}$ indicates that integration by parts is considered. See also extensions to the case of higher order susceptibilities in \cite{lucarini2005,lucarini08,L09,LC12}.

As discussed in \cite{LS11,LBHRPW14,RLL14}, the Ruelle response theory provides a powerful language for framing the problem of the response of the climate system to perturbations. 
Clearly, given a  vector flow $F(x,t)$, it is possible to define different background states, corresponding to different reference climate conditions, depending on how we break up $F(x,t)$ into the two contributions $F(x)$ and $\epsilon X(x,t)$ in the right hand side of Eq. \ref{eqpertu}. Nonetheless, as long as the expansion is well defined, the sum given in Eq. \ref{mut} does not depend on the reference state. Of course, a wise choice of the reference dynamics leads to faster convergence. 

Note that once we define a background vector flow $F(x)$ and approximate its invariant measure $\bar{\mu}(\mathrm{d}x)$ by performing an ensemble of simulations, by using Eqs. \ref{mut}-\ref{lrruelle} we can construct the  time dependent measure $\mu_t(\mathrm{d}x)$ for many different choices of the perturbation field $\epsilon X(x,t)$, as long as we are within the radius of convergence of the response theory. Instead, in order to construct the  time dependent measure following directly the definition of the  pullback attractor, we need to construct a different ensemble of simulations for each choice of $F(x,t)$. 

One needs to note that constructing directly the response operator using the Ruelle formula given in Eq. \ref{lrruelle} is indeed challenging, because of the different difficulties associated to the contribution coming from the unstable and stable directions \cite{AM07a}; nonetheless, recent applications of adjoint approaches \cite{EHL04} seem quite promising \cite{W13}. 

Instead, starting from Eqs. \ref{lrruelle}-\ref{lrruelle2}, it is possible  provide a simple yet general method for predicting the response the system for any observable $\Psi$ at any finite or infinite time horizon $t$ for any time modulation $T(t)$ of the vector field $X(x)$, if the corresponding Green function or, equivalently, the susceptibility, is known. Moreover, given a specific choice of $T(t)$ and measuring the $\langle \Psi\rangle_0^{(1)}(t)$ from a set of experiments, the same equations allow one to derive the Green function. Therefore, using the output of a specific set of experiments, we achieve predictive power for \textit{any} temporal pattern of the forcing $X(x)$. In other term, from the knowledge of the time dependent measure of one specific pullback attractor, we can derive the time dependent measures of a family pullback attractors. We will follow this approach in the analysis detailed below. While the methodology is almost trivial in the linear case, it is in principle feasible also when higher order corrections are considered, as long as the response theory is applicable \cite{lucarini08,L09,LC12}. 

We also wish to remark that in some cases divergence in the response of a chaotic system can be associated to the presence of slow decorrelation for the measurable observable in the background state, which, as discussed in \textit{e.g.} \cite{chekroun2014}, can be related to the presence of nearby critical transitions. Indeed, we have recently investigated such issue in \cite{Tantet2015b}, thus providing the statistical mechanical analysis of the classical problem of multistability of the Earth's system previously studied using macroscopic thermodynamics in \cite{Luchyst,Boschi,Lucarini2013a}.

%\newpage

\section{A Climate Model of Intermediate Complexity: the Planet Simulator - PLASIM}\label{methods}
The Planet Simulator (PLASIM) is a climate model of intermediate
complexity, freely available upon request to the group of Theoretical Meteorology at the University of Hamburg (\texttt{https://www.mi.uni-hamburg.de/en/arbeitsgruppen/theoretische-meteorologie.html}) and  includes a graphical user interface facilitating its use. By intermediate complexity we mean that the model is gauged in such a way to be parsimonious in terms of computational cost and flexible in terms of possibility to explore widely differing climatic regimes \cite{lucarini_modelling_2013}. Therefore, the most important climatic processes are indeed represented, and the model is  complex 
enough to feature essential characteristics of high-dimensional, dissipative, and chaotic systems, 
as the existence of a limited horizon of predictability due to the presence of instabilities in the flow. Nonetheless, one has to sacrifice the possibility of using the most advanced parametrizations for subscale processes and cannot use high resolutions for the vertical and horizontal directions in the representation of the geophysical fluids. Therefore, we are talking of a modelling strategy that differs from the conventional approach aiming at achieving the highest possible resolution in the fluid fields and the highest precision in the parametrization of the highest possible variety of subgrid scale processes  \cite{IPCC13}, but rather focuses on trying to reduce the gap between the modelling and the understanding of the dynamics of the geophysical flow \cite{Held05}.

The dynamical core of PLASIM is based on the Portable University Model of the Atmosphere PUMA  
\cite{Fraedrich1998}. The atmospheric dynamics is modelled using the primitive equations formulated 
for vorticity, divergence, temperature and the logarithm of surface pressure. Moisture is included by 
transport of water vapour (specific humidity). The governing  equations are solved using the spectral 
transform method \cite{Eliasen1970, Orszag1970}. In the vertical, non-equally spaced sigma (pressure 
divided by surface pressure) levels are used. The parametrization of unresolved processes 
consists of long- \cite{Sasamori1968} and short-  \cite{Lacis1974} wave radiation, interactive clouds 
\cite{Stephens1978, Stephens1984, Slingo1991}, moist  \cite{Kuo1965, Kuo1974} and dry convection, 
large-scale precipitation, boundary layer fluxes of latent and sensible heat and vertical and horizontal diffusion
\cite{Louis1979, Louis1981, Laursen1989, Roeckner1992}. The land surface scheme uses five diffusive 
layers for the temperature and a bucket model for the soil hydrology. The oceanic part is a 50 m 
mixed-layer (swamp) ocean, which includes a thermodynamic sea ice model  \cite{Semtner1976}. 

The horizontal transport of heat in the ocean can either be prescribed or parametrized by horizontal 
diffusion. In this case, we consider the second setting, as opposed to what explored in \cite{RLL14}, because it is well known that having even a severely simplified representation of the large scale heat transport performed by the ocean improves substantially the realism of the resulting climate. We remind that the ocean contributes to about 30\% of the total meridional heat transport in the present climate \cite{Peixoto:1992,Fas2,Lucarini:2011_RG}. A detailed study of the impact of changing oceanic heat transports on the dynamics and thermodynamics of the atmosphere can be found in \cite{Knietzsch2015}.

The model is run at T21 resolution (approximately $5.6^o\times5.6^o$)
with 10 vertical levels. While this resolution is relatively low, it
is expected to be sufficient for obtaining a reasonable description
of the large scale properties of the atmospheric dynamics, which
are most relevant for the global features we are interested in. We
remark that previous analyses have shown that using a spatial resolution
approximately equivalent to T21 allows for obtaining an
accurate representation of the major large scale features of the climate
system. PLASIM features $O(10^5)$ degrees of freedom, while state-of-the-art Earth System Models boast easily over $10^8$ degrees of freedom. 

\begin{figure}
\centering{a) \includegraphics[width=0.45\textwidth]{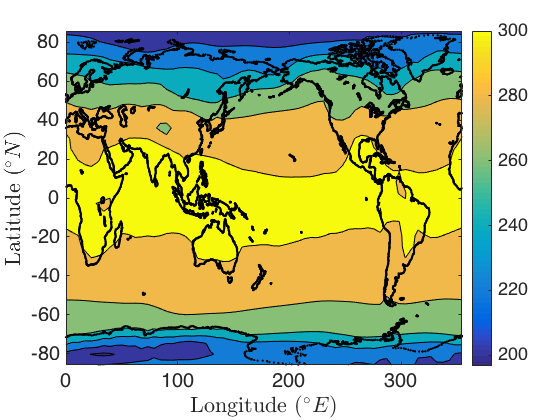}
b) \includegraphics[width=0.45\textwidth]{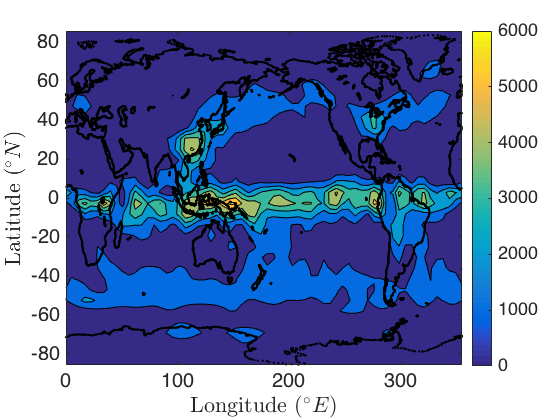}\\
c) \includegraphics[width=0.45\textwidth]{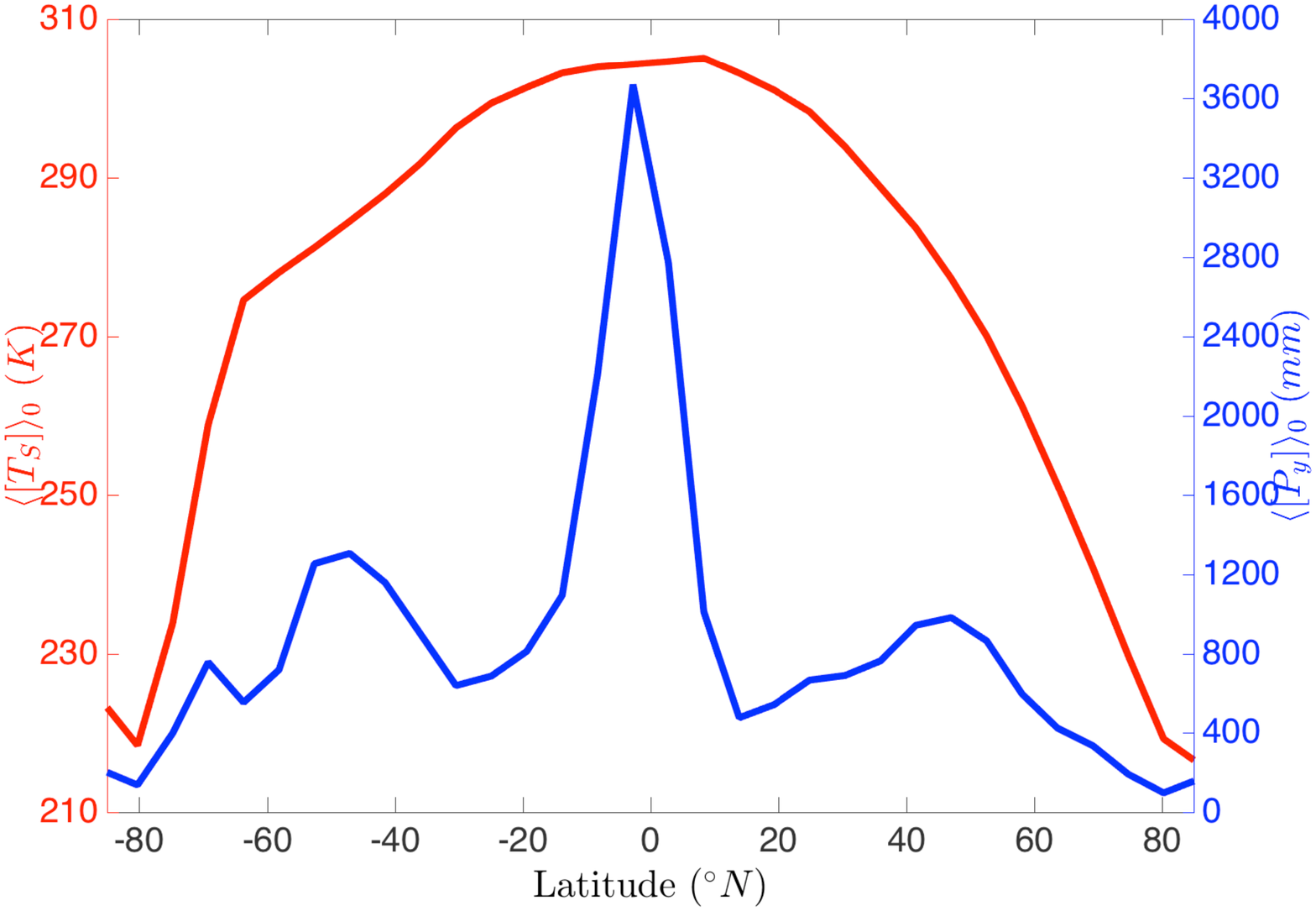}}
\caption{Long-term climatology of the PLASIM model. Control run performed with background value of $CO_2$ concentration set to 360 ppm and solar constant defined as  $S=1365$ $Wm^{-2}$: a) Surface temperature $\langle T_S \rangle_0 $ (in $K$); b) Yearly total precipitation $\langle P_y \rangle_0 $ (in $mm$); c) Zonally averaged values $\langle [T_S] \rangle_0 $ (red line and red $y-$axis) and $\langle [P_y]\rangle_0$  (blue line and blue $y-$axis). }\label{Fig1}
\end{figure}

While missing a dynamical ocean hinders the possibility of having a good representation of the climate variability on multidecadal or longer timescales, the climate  simulated by PLASIM is definitely Earth-like, featuring qualitatively correct large scale features 
and turbulent atmospheric dynamics. Figure \ref{Fig1} provides an outlook of the climatology of the model run with constant $CO_2$ concentration of $360$ $ppm$ and solar constant set to $S=1365$ $Wm^{-2}$. We show the long-term averages of the surface temperature $T_S$ (panel a) and of the yearly total precipitation $P_y$  (panel b) fields, plus their zonal averages $[T_S]$ and $[P_y]$\footnote{We indicate with $[A]$ the zonally averaged surface value of the quantity $[A]$.}. Despite the simplifications of the model, one finds substantial agreement with the main features of the climatology obtained from observations and state-of-the-art model runs: the average temperature monotonically decreases as we move poleward, while precipitation peaks at the equator, as a result of the large scale convection corresponding to the intertropical convergence zone (ICTZ), and features two secondary maxima at the mid latitudes of the two hemispheres, corresponding to the areas of the so-called storm tracks \cite{Peixoto:1992}. As a result of the lack of a realistic oceanic heat transport and of too low resolution in the model, the position to the ICTZ is a bit unrealistic as it is shifted southwards compared to the real world, with the precipitation peaking just south of the equator instead of few degrees north of it. 

Beside standard output,  PLASIM provides  comprehensive 
diagnostics for the nonequilibrium thermodynamical properties of the climate system and in 
particular for  local and global energy and entropy budgets. PUMA and PLASIM have already  been used for several theoretical climate studies, including a variety 
of problems in  climate response theory \cite{RLL14,LBHRPW14}, climate thermodynamics  
\cite{Fraedrich2008,LucACP}, analysis of climatic tipping points \cite{Luchyst,Tantet2015b}, and in the dynamics of exoplanets  
\cite{Boschi,Lucarini2013a}. 

\subsection{Experimental Setting}\label{procedure}
We want to perform predictions on the climatic impact of different scenarios of increase in the $CO_2$ concentration with respect to a baseline value of $360$ $ppm$, focusing on observables $\Psi$ of obvious climatic interest such as, \textit{e.g.} the globally averaged surface temperature $T_S$. We wish to emphasise that  most state-of-the-art general circulation
models feature an imperfect closure of the global energy budget
of the order of $1$ $Wm^{-2}$ for standard climate conditions, due to
inaccuracies in the treatment of the dissipation of kinetic energy
and the hydrological cycle \cite{Lucarini:2011_RG,LBHRPW14,Liepert,Liepert2}. Instead, PLASIM has been modified in such a way that a more accurate representation of the energy budget is obtained, even in rather extreme climatic conditions \cite{Luchyst,Boschi,Lucarini2013a}. Therefore, we are confident of the thermodynamic consistency of our model, which is crucial for evaluating correctly the climate response to radiative forcing resulting from changes in the opacity of the atmosphere.

We proceed step-by-step as follows:
\begin{itemize}
\item We take as dynamical system $\dot{x}=F(x)$ the spatially discretized version of the partial differential equations describing the evolution of the climate variables in a baseline scenario with set boundary conditions and set values for, \textit{e.g.},  the $CO_2$ $=$ $360$ $ppm$ baseline concentration and the solar constant $S=1365$ $Wm^{-2}$. We assume, for simplicity, that system does not feature daily or seasonal variations in the radiative input at the top of the atmosphere. We run the model for 2400 years in order to construct a long control run. Note that the model relaxes to its attractor with an approximate time scale of 20-30 years.  
%\item Let us choose for the observable $A$ the globally averaged surface temperature of the planet $T_S$. 
\item We study the impact of perturbations using a specific test case. We run a first set of $N=200$ perturbed simulations, each lasting 200 years, and each initialized with the state of the model at year $200 + 10k$, $k=1,\ldots,200$. We choose as  perturbation field $X(x)$ the additional convergence of radiative fluxes due to changes in the atmospheric $CO_2$ concentration. Therefore, such a perturbation field has non-zero components only for the variables directly affected by such forcings, \textit{i.e.} the values of the temperatures at the resolved grid points of the atmosphere and at the surface.   In each of these simulation we perturbed the vector flow by doubling instantaneously the $CO_2$ concentration. This corresponds to  having $\dot{x}=F(x)\rightarrow \dot{x}=F(x)+\epsilon \Theta(t)X(x)$. Note that the forcing is well known to scale proportionally with to the logarithm of the $CO_2$ concentration \cite{Peixoto:1992}. 
\item By plugging $T(t)=T_a(t)=\Theta(t)$ into Eqs. \ref{lrruelle}, we have that :
\begin{equation}
\frac{\mathrm{d}}{\mathrm{d}t}\langle \Psi\rangle_0^{(1)}(t)=\epsilon G_\Psi^{(1)}(t)\label{computeG}
\end{equation}
 We estimate $\langle \Psi\rangle_0^{(1)}(t)$ by taking the average of response of the system over a possibly large number of ensemble members, and use the previous equation to derive our estimate of $ G_\Psi^{(1)}(t)$, by assuming linearity in the response of the system. In what follows, we present the results obtained using all the available $N=200$ ensemble members, plus, in some selected cases, showing the impact of having a smaller number ($N=20$) of ensemble members. 

It is important to emphasize that framing the problem of climate change using the formalism of response theory gives us  ways for providing simple yet useful formulas for defining precisely the climate sensitivity $\Delta_\Psi$  for a general observable $\Psi$, as $\Delta_\Psi=\Re\{\chi^{(1)}_{\Psi}(0)\}$. Furthermore, if we consider perturbation modulated by a Heaviside distribution, we have the additional simple and useful relation:
\begin{equation}
\Delta_\Psi=\frac{2}{\pi \epsilon }\int d\omega'\Re\{\langle \Psi \rangle_0^{(1)}{(\omega')}\},
\end{equation}
which relates climate response at all frequencies to its sensitivity, as resulting from the validity of the Kramers-Kronig relations. 

We remark that using a given set of forced experiments it is possible to derive information on the climate response to the given forcing for as many climatic observables as desired. It is important to note that, for a given finite intensity $\epsilon$ of the forcing, the accuracy of the linear theory in describing the full response depends also on the observable of interest. Moreover, the signal to noise ratio and, consequently, the time scales over which predictive skill is good may change a lot from variable to variable.

\item We want to be able to predict at finite and infinite time the response of the system to some other pattern of $CO_2$ forcing. Following \cite{RLL14}, we choose as a pattern of forcing one of the classic IPCC scenarios, namely a 1\% yearly increase of the $CO_2$ concentration up to its doubling, and we perform a set of additional $N=200$ perturbed simulations performed according to such a protocol. This corresponds to choosing a new time modulation that can be approximated as a linear ramp of the form 
\begin{equation}
T_b^\tau(t)=\epsilon t/\tau, \quad 0\leq t \leq \tau, \quad T_b^\tau(t)=1, t>\tau,\label{ramp}
\end{equation}
where $\tau=100 \log 2$ years $\sim 70$ years is the doubling time. Therefore, for each observable $\Psi$ we compare the result of convoluting the estimate of the Green function obtained in the previous step with time modulation $T_b(t)$ with the ensemble average obtained from the new set of simulations.   
\end{itemize}

%Apparently, despite all the nonlinear feedbacks of the climate model, the response to changes in the logarithm of $CO_2$ concentration can be accurately described by linear response theory at all time scales. Nonlinearity in the underlying equations and presence of strong positive and negative feedbacks do not rule out the possibility of constructing accurate methods for predicting the response. In fact, the methods described here could be extended to the nonlinear case by looking at the response in the frequency domain \citep{lucarini08,Lucarini:2009_JSP}, even if the data quality requirement is obviously stricter.

%The result presented here suggests that many of the scenarios of greenhouse gases concentration included in the IPCC reports \cite{IPCC01,IPCC07,IPCC13} may in fact be partly redundant, as for certain variables  might be accurately described by linear response theory starting from just one scenario.  Equations \ref{risposta3} - \ref{rispostanew} constitute the basis for predicting climate response at all scales. 

\section{Results}\label{results}

The response theory sketched above allows us in principle to study the change in the statistical properties of any well-behaved, smooth enough observable. Nonetheless, problems naturally emerge when we consider finite time statistics, finite number of ensemble members, and finite precision approximations of the response operators. %, because different observables can be more or less noisy, and the reconstruction of the corresponding Green function can in some cases be problematic. The issue is complicated by the fact that, depending on the signal-to-noise ratio, we are constrained to studying the response for suitably time average quantities, see \cite{RLL14} for a thorough discussion of this important point.
{ The Green functions of interest are derived using Eq. \ref{computeG} as time derivative of the ensemble averaged time series of the observed response to the probe forcing whose time modulation is given by the Heaviside distribution. Clearly, the response is not smooth unless the ensemble size $N\rightarrow \infty$. Therefore, taking numerically the time derivative leads to having a very noisy estimate of the Green function, which  might also depend heavily on the specific procedure used for computing the discrete derivative. This might suggest that the procedure is not robust. Instead, we need to keep in mind that we aim at using the Green function \textit{exclusively as a tool for predicting the climate response}. Therefore, if we convolute with the Green function with a sufficiently smooth modulating factor $T(t)$ as in Eq. \ref{lrruelle}, the small time scales fluctuations of the Green function, albeit large in size, become of no relevance, because they are averaged out. This is further eased if, instead of looking for predictions valid for observables defined at the highest possible time resolution of ur model, we concentrate of suitably time averaged quantities. Clearly, while it is in principle possible to define mathematically the climate response on the time scale of, \textit{e.g.}, one second, this has no real physical relevance. Looking at the asymptotic behaviour of the susceptibility it is possible to derive what is, depending on the signal-to-noise ratio, the time scale over which we can expect to be able to perform meaningful predictions; see discussion in \cite{RLL14}.}

\begin{figure}[ht]
a) \includegraphics[width=0.47\textwidth]{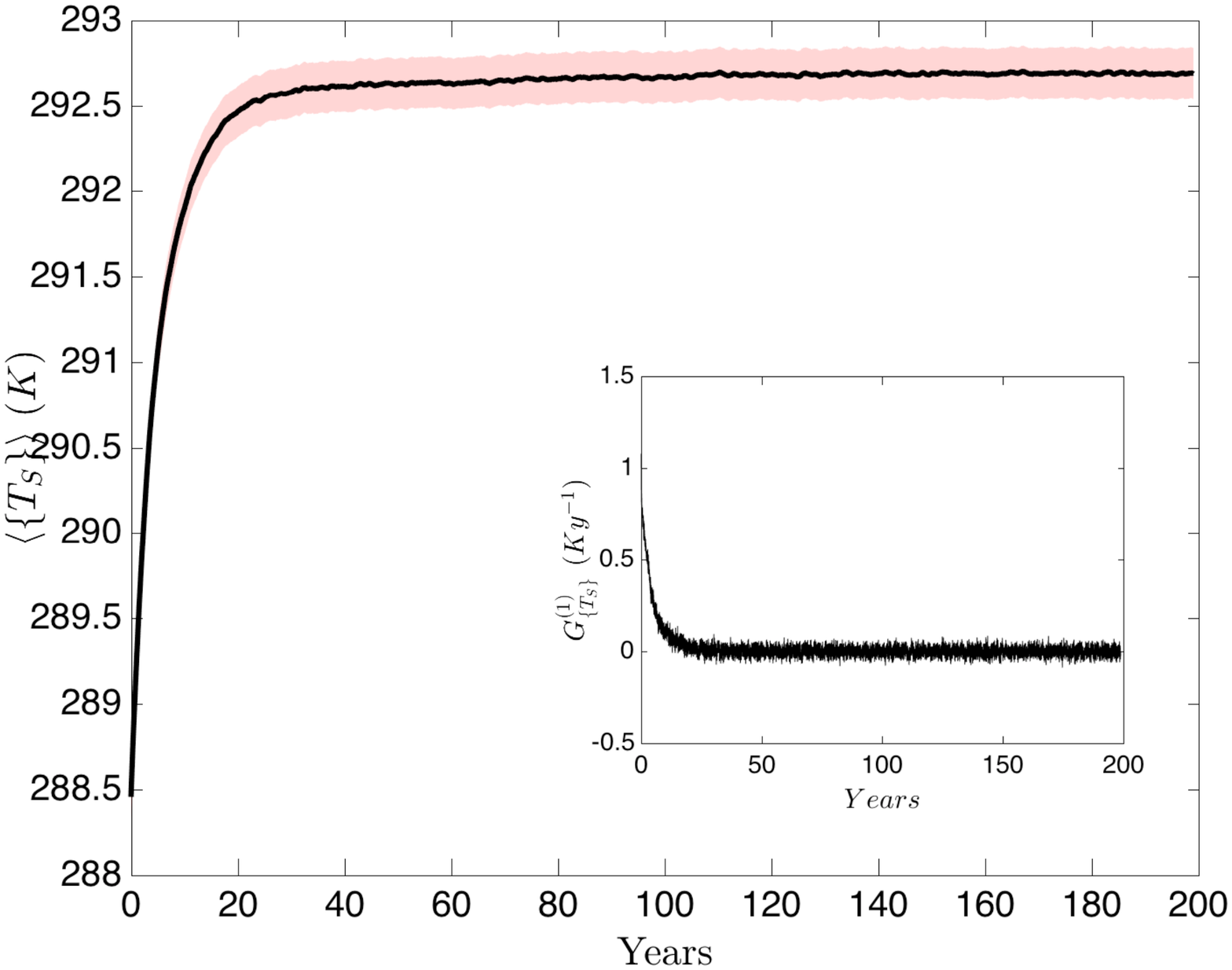}
b) \includegraphics[width=0.47\textwidth]{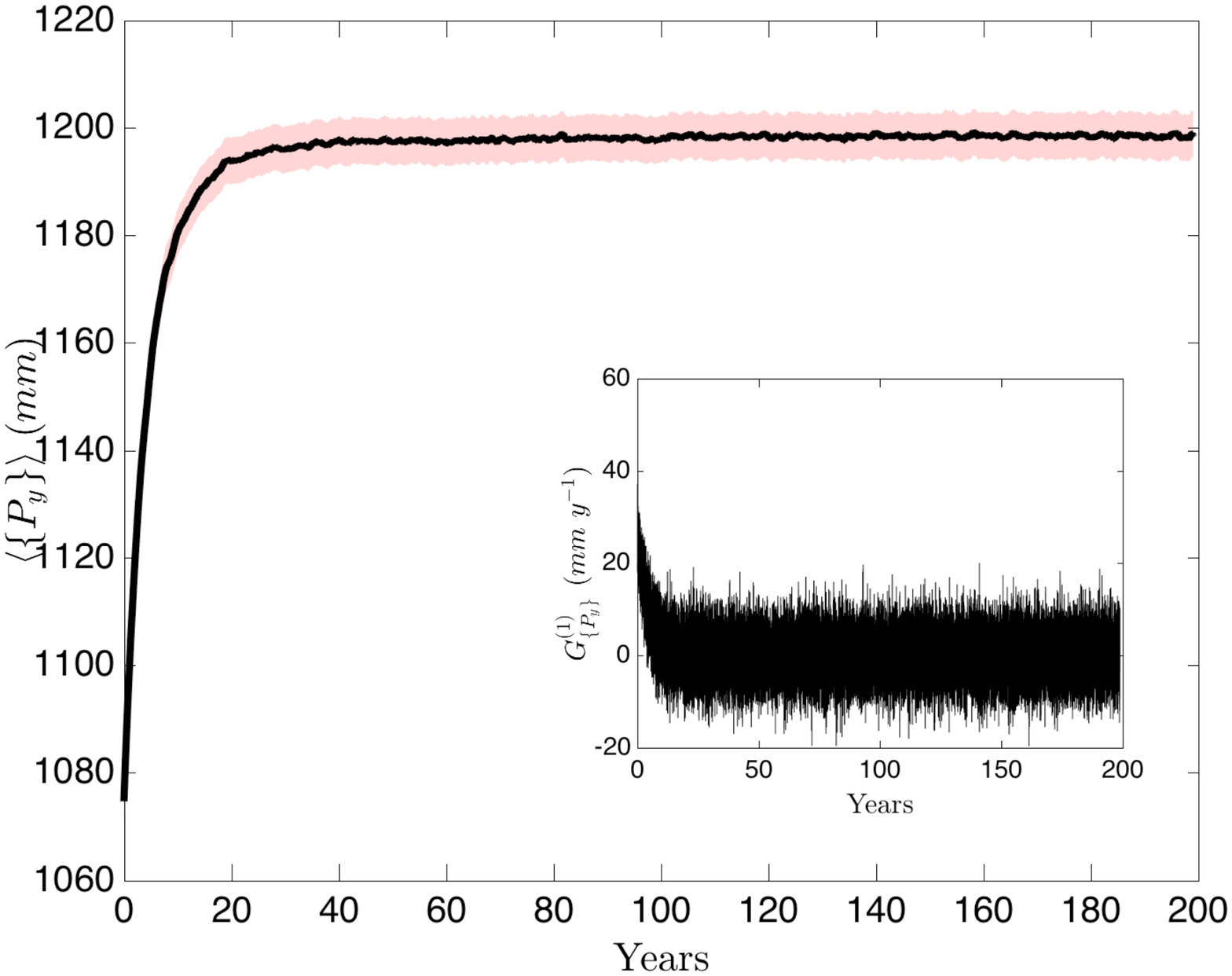}\\
\caption{Climate response to instantaneous $[CO_2]$ doubling for (a) globally averaged mean annual surface temperature $\langle\{ T_S\} \rangle$; and (b) the globally averaged annual total precipitation $\langle \{P_y\}\rangle$ (b). The black line shows the $N=200$ ensemble averaged properties of the doubling $CO_2$ experiments. The light red shading indicates the variability as given by the $2\sigma$ of the ensemble members. In each panel, the inset portrays the corresponding Green function.}\label{Fig2}
\end{figure}

In order to provide an overlook of the practical potential of the response theory in addressing the problem of climate change, we have decided to focus on two  climatological quantities of general interest, namely the yearly averaged surface temperature $T_S$ and the yearly total precipitation $P_y$. Such quantities have obvious relevance for basically any possible impact study of climate change, and, while there is much more in climate change than studying the change in $T_S$ and $P_y$, these are indeed the first quantities one considers when benchmarking the performance of a climate model and when assessing whether climate change signals can be detected.

Another issue one needs to address is the role of the spatial patterns of change in the considered quantities. The change in the globally average surface temperature $\{ T_S\}$\footnote{We indicate with $\{A \}$ the globally averaged surface value of the quantity $A$.} has undoubtedly gained prominence in the climate change debate and in the IPCC negotiations targets are tailored according to such an indicators. Nonetheless, the impacts of climate change are in fact local and one needs to investigate the geography of the change signals \cite{IPCC13}. Evidently, one expects that coarse grained (in space) quantities will have a better signal-to-noise ratio and will allow for performing higher precision climate projection using response theory. On the other side, the performance of linear response theory at local scale might be hindered by the presence of local strongly nonlinear feedbacks, such as the ice-albedo feedback, which have less relevance when spatial averaging is performed. In what follows, we will consider observables constructed from the spatial fields of $T_S$ and $P_y$ by performing different levels of coarse graining. We will begin by looking into globally averaged quantities, and then address the problem of predicting spatial pattens of climate change.

\subsection{Globally Averaged Quantities}

We begin our investigation by focusing on the globally averaged surface temperature $\{T_S\}$ and the globally averaged yearly total precipitation $ \{P_y\} $. In what follows, we perform the analysis using the model output at the highest available resolution (1 day) but present, for sake of convenience  and since we indeed focus on yearly quantities, data where a $1-$year band pass filtering is performed. 

\subsubsection{From Green Functions to Climate Predictions}

Figure \ref{Fig2} shows the ensemble average performed over $N=200$ members of the change of $\langle \{T_S\}\rangle$ (panel a) and $\langle \{P_y\} \rangle$ (panel b)  as a result of the instantaneous doubling of the  $CO_2$ concentration described in the previous section. We find that the asymptotic change in the surface temperature is given by the equilibrium climate sensitivity $ECS=\Delta_{\{T_S\}}\sim 4.9$ $K$, which is just outside the \textit{likely} range of values for the ECS elicited in \cite{IPCC13}. Note that the models discussed in \cite{IPCC13} include more complex physical and chemical processes and most notably a comprehensive representation of the dynamics of the ocean, plus featuring a seasonal and daily cycle of radiation, so that the comparison in a bit unfair. Nonetheless, we get the sense that PLASIM features an overall reasonable response to changes in the $CO_2$. This is confirmed  by looking at the long terms response of $\langle \{P_y\} \rangle$ to $[CO_2]$ doubling, for which we find $\Delta_{\{P_y\}}\sim 125$ $mm$, which corresponds to about $11.6\%$ of the initial value. These figures are also in good agreement with what reported in \cite{IPCC13}. We will comment below on the relationship between the climate change signal for $\{T_S\}$ and for $\{P_y\}$.  

In each panel of Fig. \ref{Fig2} we show as inset the corresponding Green function computed according to Eq. \ref{computeG}. We find that both Green functions have to first approximation an exponential behaviour, even if one can expect also important deviations, as discussed in \cite{RLL14}. We will not elaborate on this. Instead we note that  $G_{\{P_y\}}^{(1)}$ is  more noisy that $G_{\{T_S\}}^{(1)}$, as a result of the fact that $\langle \{P_y\} \rangle^{(1)}(t)$ has stronger high frequency contribution to its variability than $\langle \{T_S\} \rangle^{(1)}(t)$, \textit{i.e.} $\langle \{P_y\} \rangle^{(1)}(t)$ has, unsurprisingly, has a much shorter decorrelation time, because it refers to the much faster hydrological cycle related processes.

\begin{figure}[ht]
a) \includegraphics[width=0.5\textwidth]{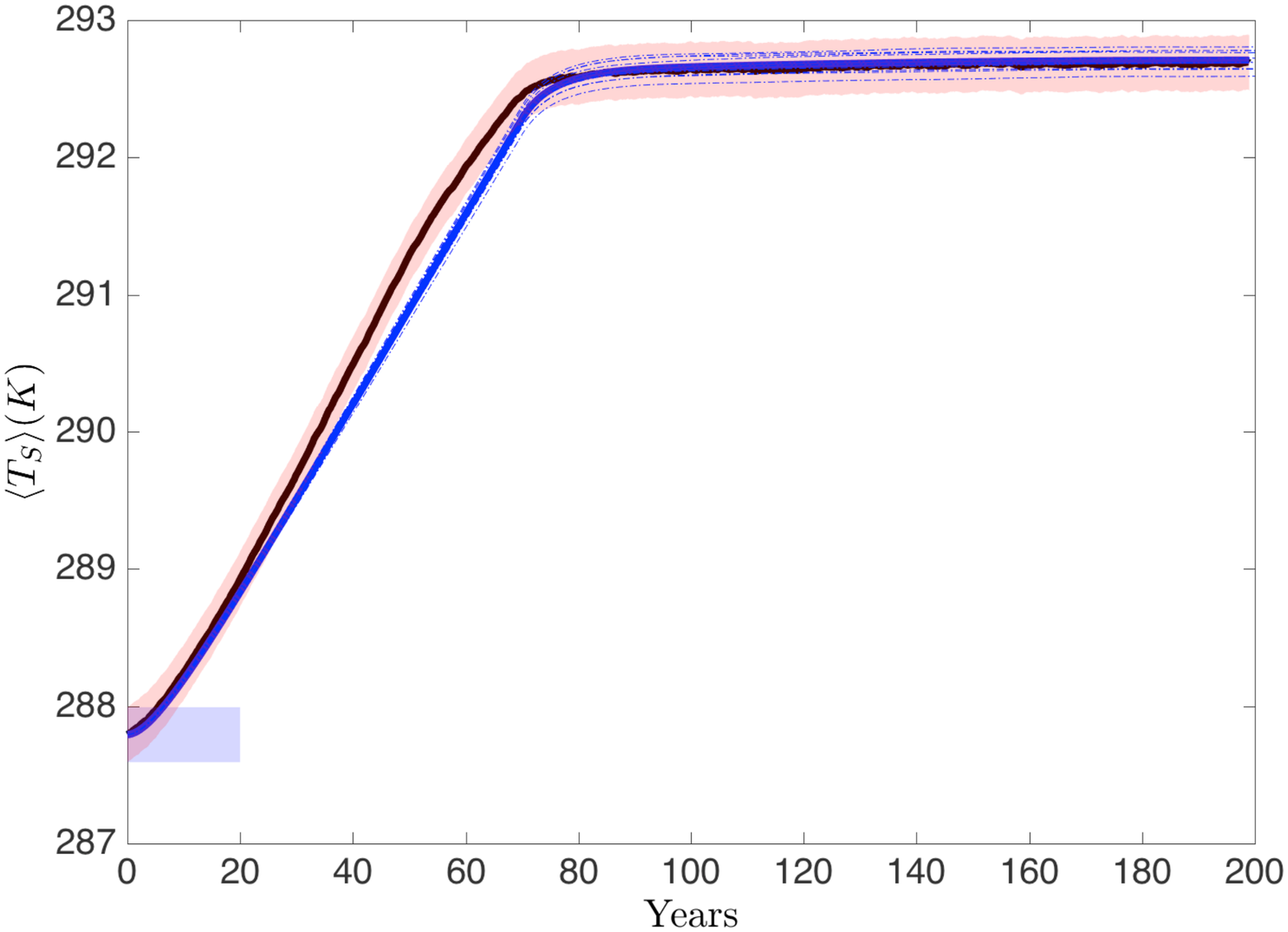}
b) \includegraphics[width=0.47\textwidth]{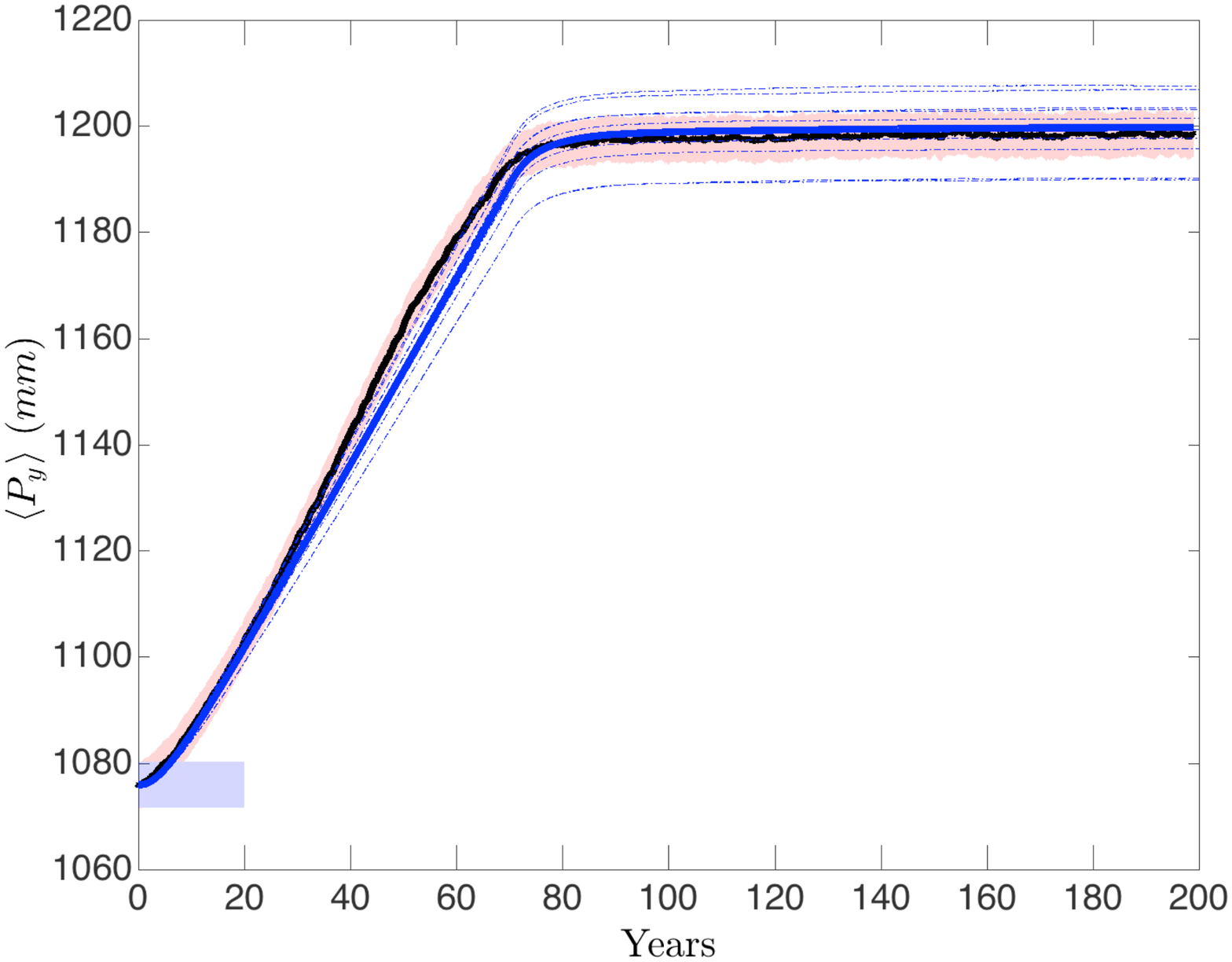}\\
\caption{Climate change projections for the globally averaged mean annual surface temperature $\langle \{T_S\} \rangle$; and (b) the globally averaged annual total precipitation $\langle \{P_y\}\rangle$. The black line shows the $N=200$ ensemble averaged properties of the experiments where we have a $1\%$ per year $[CO_2]$ increase up to $[CO_2]$ doubling. The light red shading indicates the variability as given by the $2\sigma$ of the ensemble members. The blue shading indicates the interannual variability of the control run. The thick blue line is the projection obtained using the Green functions derived using $N=200$ ensemble members of the instantaneous doubling $[CO_2]$ experiments. The dashed lines correspond to ten projections each obtained using Green functions derived from $N=20$ ensemble members of the instantaneous doubling $[CO_2]$ experiments.}\label{Fig3}
\end{figure}

Figure \ref{Fig3} provides a comparison between the statistics of $\langle \{T_S\} \rangle(t)$ and $\langle \{P_y\} \rangle(t)$ obtained by performing $N=200$ simulations where we increase the $CO_2$ concentration by $1\%$ per year until doubling, and the prediction of the response theory derived by performing the convolution of the Green functions shown in Fig. \ref{Fig2} with the ramp function given in Eq. \ref{ramp}. We have that the prediction of the ensemble average $\delta\langle  \{T_S\}  \rangle_0^{(1)}$ and  of the ensemble average $\delta\langle  \{P_y\}  \rangle_0^{(1)}$ (blue thick lines) obtained using $N=200$ ensemble members for the doubling $CO_2$ experiments is in good agreement for both observables with the results of the direct numerical integrations.

More precisely, we have that the prediction given by the climate response lies within the variability of the $N=200$ direct simulations for basically all time horizons. The range of variability is depicted with a light red shade, centered on the ensemble mean represented by the black line. Instead, within a time window of about 40 to 60 years, the response theory slightly underestimates the true change in both $\{T_S\}$ and $\{P_Y\}$: we will investigate below the reasons for this mismatch.

It is well known that it is hard to construct a very large set of ensemble members in the case of state-of-the-art climate models, due to the exorbitant computing costs associated with each individual run. Additionally, in general, in the modelling practise the continuous growth in computing and storing resources is typically invested in increasing the resolution and the complexity of climate models, rather than populating more attentively the statistics of the run of a given version of a model. Therefore, even in coordinated modelling exercises contributing to the latest IPCC report, the various modelling groups are requested to deliver a number of ensemble member of the order of 10 \cite{IPCC13}. 

In order to partially address the problem of assessing how the number of ensemble members can affect our   prediction, we present in Fig. \ref{Fig3} the result of the prediction of climate change signal for $\langle \{T_S\} \rangle$ and $\langle \{P_y\} \rangle$ obtained by constructing the Green function using only $N=20$ members of the ensemble of simulations of instantaneous $CO_2$ concentration doubling. Ten thin dashed blue lines represent in each panel the result of such predictions. Interestingly, each of the prediction obtained with a reduced number of ensemble members also  agrees with the direct numerical simulations when we consider  $\langle \{T_S\} \rangle$, because the spread around our best estimates obtained using the full set of ensemble members is minimal. 

Instead, when looking at $\langle \{P_y\} \rangle$, we have that only some of the predictions derived using  reduced ensemble sets lie within the variability of the direct numerical simulations, with a much larger spread around the prediction obtained with the full ensemble set. This fact is closely related to the fact that the corresponding Green function is noisier, see Fig. \ref{Fig2}, and suggests that in order to have good convergence of the statistical properties of the response operator a better sampling of the attractor is needed.  

We conclude that the computational requirements for having good skill in predicting the changes in $\{P_y\}$ are harder than in the case of $\{T_S\}$. Indeed, the surface temperature is a \textit{good} quantity in terms of our ability to predict it, and, in terms of being a good indicator of climate change, as it  allows one to find clear evidence of the departure of the statistics from  the unperturbed  climate conditions. This is in agreement with the actual practice of the climate community \cite{IPCC13}. 

\subsubsection{Climate Change Detection and Climate Inertia}

We dedicate some additional care in studying the climate response in terms of changes in the globally averaged surface temperature. We wish to use the information gathered so far for assessing some features of climate change in different scenarios of modulation of the forcing. In particular, we focus on studying the properties of the following expression:
\begin{equation}
\delta\langle \{T_S\}\rangle_0^{(1)}(t,\tau)=\int_0^\infty \mathrm{d}s G_{\{T_S\}}^{(1)}(t-s) T_b^\tau(s)\label{projchange}
\end{equation}
when different values of the $CO_2$ concentration doubling time $\tau$ are considered. This amounts to performing a family of climate projections where the rate of increase of the $CO_2$ concentration is $r_\tau=100(2^{(1/\tau)}-1)$ $\%$ per year (where $\tau$ is expressed in years). As limiting cases, we have $\tau$=0 - instantaneous doubling, as in fact described by the \textit{probe} scenario $T_a(t)$, and $\tau\rightarrow \infty$, which provides the adiabatic limit of infinitesimally slow changes.

\begin{figure}[ht]
a) \includegraphics[width=0.5\textwidth]{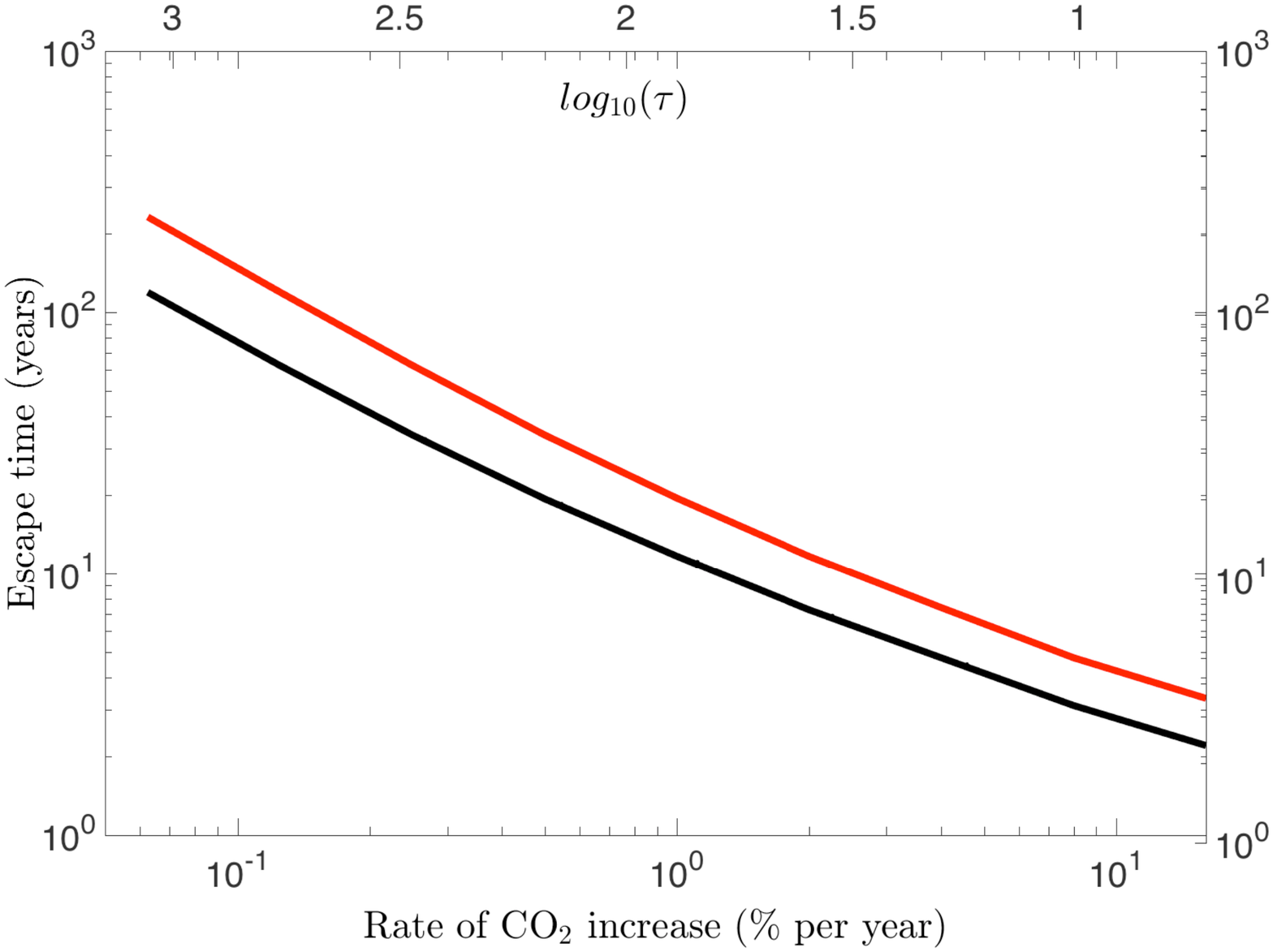}
b) \includegraphics[width=0.47\textwidth]{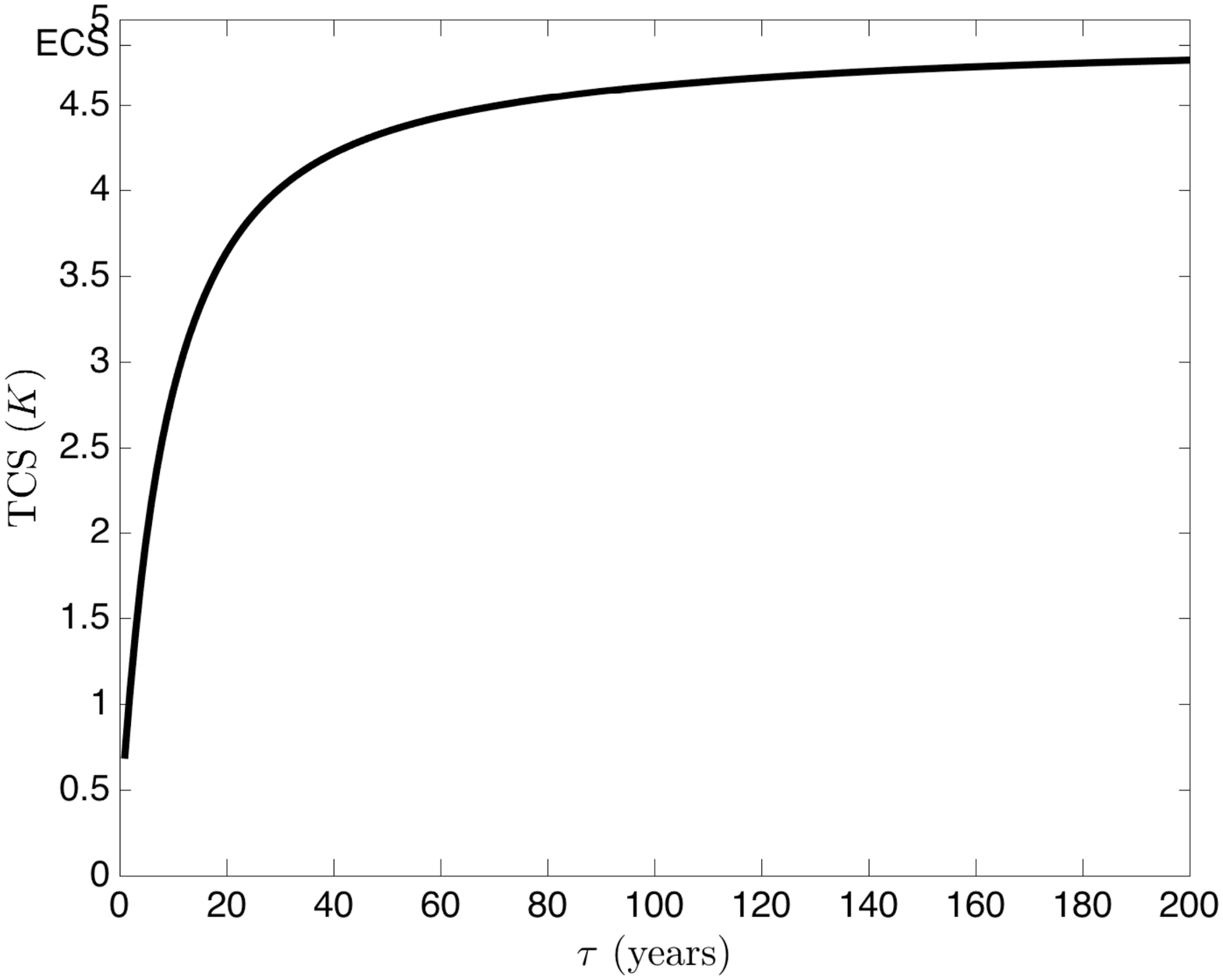}
\caption{Climate response at different time horizons. a) Time needed for detecting climate climate change: ensemble average of the response is outside the interannual variability with $95\%$ statistical significance (black line); no overlap between the $95\%$ confidence intervals representing the interannual variability of the control run and the ensemble variability of the projection (red line). b) Transient climate sensitivity - TCS - as measured by the ensemble average of the $\langle T_S \rangle$ at time $\tau$ where doubling of $[CO_2]$ is realized following an exponential increase at rate of $r_\tau=100(2^{(1/\tau)}-1)$ $\%$ per year, where $\tau$ is expressed in years. The equilibrium climate sensitivity (ECS) is indicated.}\label{Fig4}
\end{figure}

{ We want to show how response theory - and, in particular, Eq. \ref{projchange} - can be used for providing a flexible tool in the problem of climate change detection. The definition of a suitable probabilistic framework for assessing whether an observed climate fluctuations is caused by a specific forcing is extremely relevant (including for legal and political reasons) and is since the early 2000s the subject of an intense debate \cite{Allen2003,Hannart2016}. In this case, given our overall goals, we provide a rather unsophisticated treatment of the problem.} In Figure \ref{Fig4} a) we present our results, where different scenarios of forcings are considered. The black line tells us what is the time it takes for the projected change in the ensemble average to lie outside the $95\%$ confidence interval of the statistics of the unperturbed control run, \textit{i.e.} practically being outside its range of interannual variability. More precisely, the black line portrays the following escape time:
\begin{equation}
t_{min,2}^\tau=\min_t |  \delta\langle \{T_S\}\rangle_0^{(1)}(t,\tau)\geq 2\sigma(\{T_S\})_0, 
\end{equation}
where $\sigma(\{T_S\})_0\sim 0.24$ $K$ is the standard deviation of the yearly averaged time series of $\{T_S\}$ in the control run.

Nonetheless, we would like to be able to assess when not only the projected change in the ensemble average is distinguishable from the statistics of the control run, but, rather, when a an actual individual simulation is incompatible with the statistics of the unperturbed climate, because we live in one of such realizations, and not on any averaged quantity. Obviously, in order to assess this, one would require performing an ensemble of direct simulations, thus giving no scope to any application of the response theory. We can observe, though, from Fig. \ref{Fig3}a), that the interannual variability of the control run and the ensemble variability of the perturbed run are rather similar (being the same if no perturbation is applied). Therefore, we heuristically assume that the two confidence intervals have the same width. The red line portrays the second escape time
\begin{equation}
t_{min,4}^\tau=\min_t |  \delta\langle \{T_S\}\rangle_0^{(1)}(t,\tau)\geq 4\sigma(\{T_S\})_0,
\end{equation}
such that for $t\geq t_{min,4}^\tau$ it is extremely unlikely that any realization of the climate change scenario due to a forcing of the form $T_b^\tau$ perturbed run has statistics compatible with that of the control run.  In other terms, $t_{min,4}^\tau$ provides a robust estimate of when detection of climate change in virtually unavoidable from a single run, while $t_{min,2}^\tau$ gives an estimate of the time horizon after which it makes sense to talk about climate change. We would like to remark that using the Green funtcions reconstructed from the reduced ensemble sets as shown in Fig. \ref{Fig3}, one obtains virtually indistinguishable estimates for  $t\geq t_{min,2}^\tau$ and $t\geq t_{min,4}^\tau$ for all values of $\tau$. This suggests that these quantities are rather robust.

We can detect two approximate scaling regimes, with changeover taking place for $r_\tau \sim 1\%$ per year.
\begin{itemize}
\item for large values of $r_\tau$ ($\geq 1\%$ per year), we have that $t_{min,2}^\tau, t_{min,4}^\tau \propto r_\tau^{-0.6}$
\item for moderate values of $r_\tau$ ($\leq 1\%$ per year), we have that $t_{min,2}^\tau, t_{min,4}^\tau \propto r_\tau^{-1}$. The latter corresponds to the quasi-adiabatic regime and one finds that, correspondingly, that $t_{min,4}^\tau, t_{min,2}^\tau \propto \tau$.
\end{itemize}

A quantity that has attracted considerable interest in the climate community is the so-called transient climate sensitivity (TCS), which, as opposed to the ECS, which looks at asymptotic temperature changes, describes the change of $\{T_S\}$ at the moment of $[CO_2]$ doubling following a $1\%$ per year increase \cite{Allen2002}. The difference between ECS and TSC gives a measure of the inertia of the climate system in reaching the asymptotic increase of $\{T_S\}$ realized with doubled $CO_2$ concentration. Using response theory, we can extend the concept of transient climate sensitivity by considering any rate of exponential increse of the $CO_2$ concentration. Using Eq. \ref{projchange}, we have that: 
\begin{equation}
TCS(\tau)=\delta\langle \{T_S\}\rangle_0^{(1)}(\tau,\tau)
\end{equation}
describes the change in the expectation value of $\{T_S\}$ at the end of the ramp of increase of $CO_2$ concentration. As suggested by the  argument proposed in \cite{Allen2002}, one expects that the TCS is a monotonically increasing function of $\tau$ (see Fig. \ref{Fig4} b), and the difference between the ECS and TCS becomes very small for large values of $\tau$,  because we enter the quasi-adiabatic regime where the change in the $CO_2$ is slower than the slowest internal time scale of the system.

\subsubsection{A Final Remark}
We would like to make a final remark of the properties of the response of the global observables $\{T_S\}$ and $\{P_y\}$.  Looking at Figs. \ref{Fig2} and \ref{Fig3}, one is unavoidably bound to observe that the temporal pattern of response of $\{T_S\}$ and $\{P_y\}$ are extremely similar. In agreement with \cite{Allen2002} (see also \cite{IPCC13}), we find that to a very good approximation the following scaling applies for all simulations: $$\frac{\delta \langle \{P_y\} \rangle_0^{(1)}(t)}{\langle \{P_y\} \rangle_{0}}\sim 0.025 \frac{\delta \langle \{ T_S\} \rangle_0^{(1)}(t)}{K},$$ where $K$ is one degree Kelvin. In other terms, the two Green functions $G_{\{T_S\}}^{(1)}$ and $G_{\{P_y\}}^{(1)}$ are, to a very good approximation, proportional to each other \textit{when yearly averages are considered}.

Note that this scaling relations does \textit{not} agree with the naive scaling imposed by the Clausius-Clapeyron relation controlling the partial pressure of saturated water vapour. In fact, were the Clausius-Clapeyron scaling correct, one would have  $$\frac{\delta \langle \{P_y\} \rangle_0^{(1)}(t)}{\langle \{P_y\} \rangle_{0}}\sim 0.07 \frac{\delta \langle \{ T_S\} \rangle_0^{(1)}(t)}{K}.$$ The reasons  why a scaling between changes in $\{T_S\}$ and $\{P_y\}$ exists at all, and why it looks like a modified version of a Clausius-Clapeyron-like law, are hotly debated in the literature \cite{Held2006,Allen2007,Gorman2015}. 

\subsection{Predicting Spatial Patterns of Climate Change}

\begin{figure}[ht]
\centering{\includegraphics[width=0.65\textwidth]{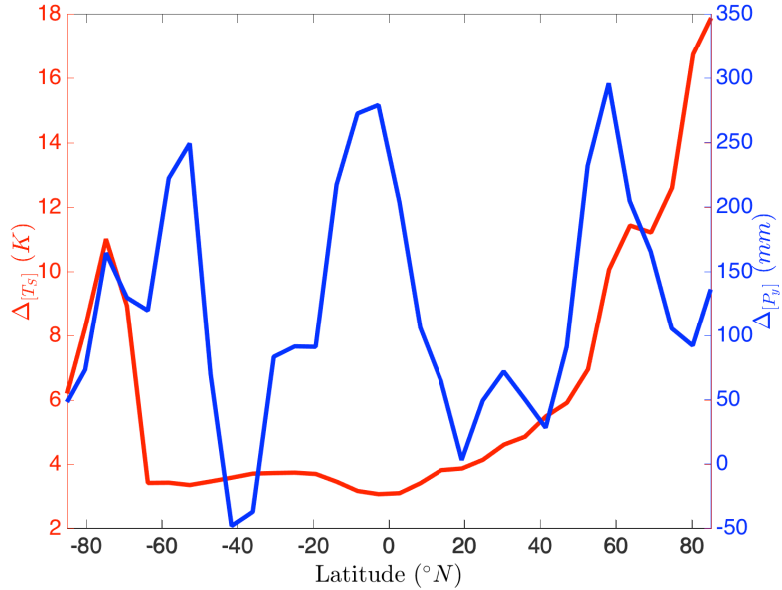}
\caption{Long-term climate change for the zonal averages of the surface temperature and of the yearly total precipitation.}\label{Fig5b}}
\end{figure}

While there is a very strong link between the change in the globally averaged precipitation and of the globally averaged surface temperature, important differences emerge when looking at the spatial patterns of change of the two fields \cite{Allen2007}. We will investigate the spatial features of climate response in the next  subsection.

\begin{figure}[ht]
a) \includegraphics[width=0.65\textwidth]{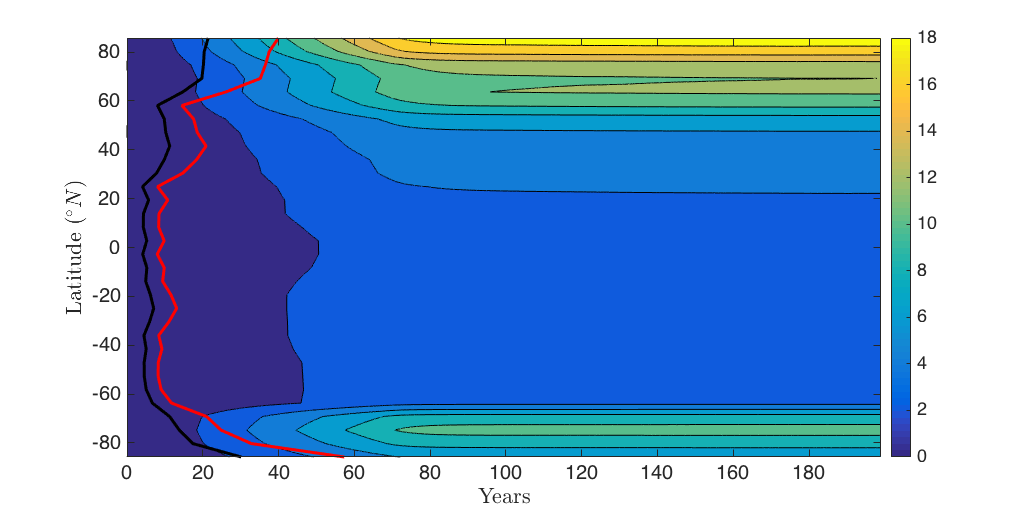}\\
b) \includegraphics[width=0.65\textwidth]{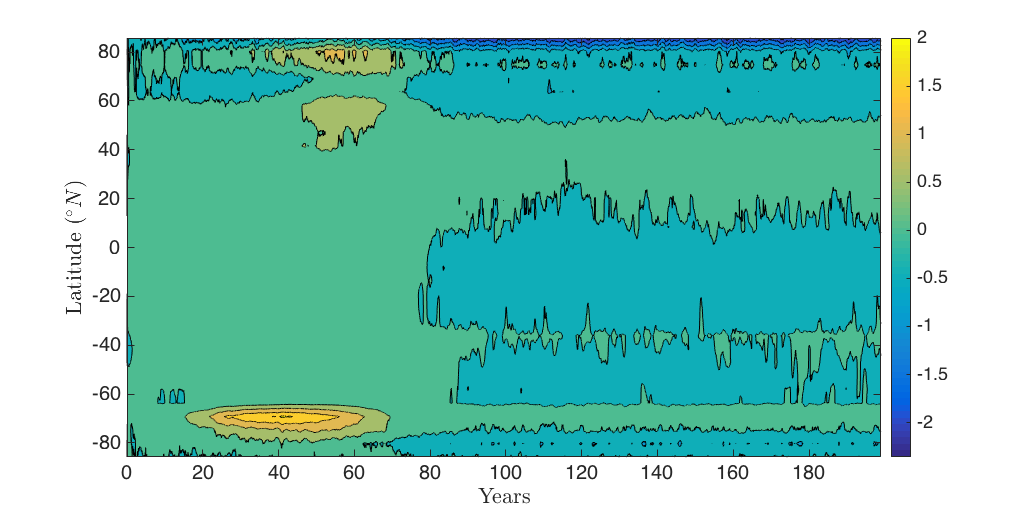}
\caption{Patterns of climate response - surface temperature $T_S$. a) Projection of the change of $[T_S]$. The black and red lines indicates the escape times as presented in Fig \ref{Fig4} for $\tau=70$ y. b) Difference between the ensemble average of the direct numerical simulations and the predictions obtained using the response theory.}\label{Fig5}
\end{figure}

The methods of response theory allow us to treat seamlessly also the problem of predicting the climate response for (spatially) local observables. It is enough to define appropriately the observable $\Psi$  and repeat the procedure described in Section \ref{procedure}. As a first step in the direction of assessing our ability to predict climate change at local scale, we mostly concentrate the zonally (longitudinally) averaged fields $[T_S](\lambda)$ and $[P_y](\lambda)$, where we have made explicit reference to to the dependence on the latitude $\lambda$. Studying these fields is extremely relevant because it allows us to look at the difference of the climate response at different latitudinal belts, which are well known to have entirely different dynamical properties, and, in particular, to look at equatorial-polar contrasts. 

\begin{figure}[ht]
a) \includegraphics[width=0.43\textwidth]{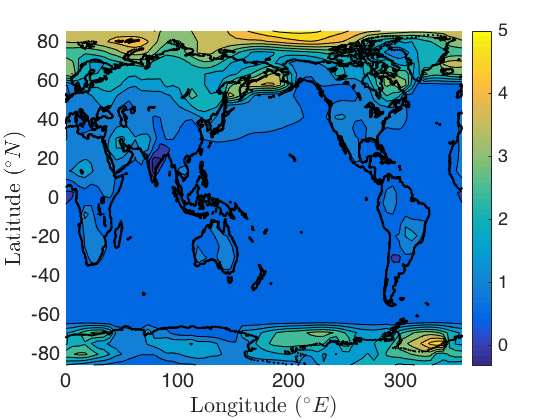}
b) \includegraphics[width=0.43\textwidth]{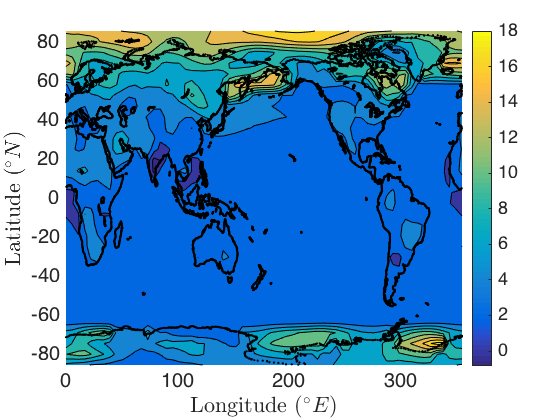}\\
c) \includegraphics[width=0.43\textwidth]{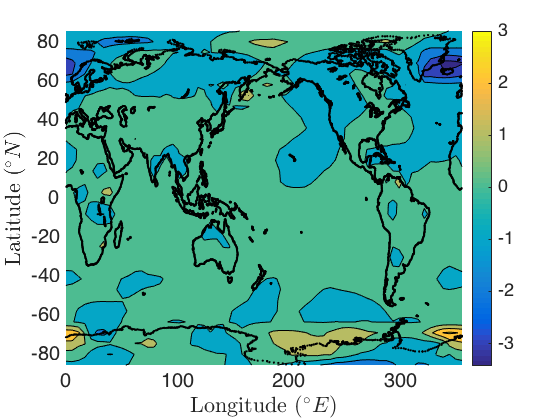}
d) \includegraphics[width=0.43\textwidth]{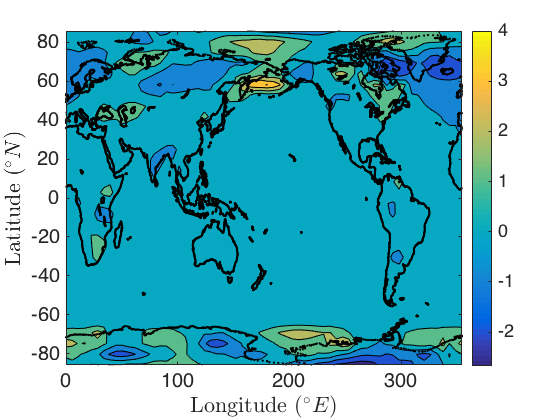}\\
\caption{Climate response at different time horizons for the $T_S$ spatial field. a) Projection obtained using response theory for a time horizon of 20 $y$. b) Same as in a), but for 60 $y$. c) Difference between the ensemble average of the direct numerical simulations and the prediction of the response theory for a time horizon of 20 $y$. d) Same as in c), but for 60 $y$.}\label{Fig6}
\end{figure}

We show in Fig. \ref{Fig5b} the long-term change in the climatology of the $[T_S]$, \textit{i.e.}, the climate sensitivity for each latitudinal band. We have confirmation of well-established findings: the response of the surface temperature is much stronger in the higher latitudes than in the tropical regions, as a result of the local ice-albedo feedback and, secondarily, of the increased transport resulting from changes in the circulation. Additionally, there is a clear asymmetry between the two hemispheres, with the response in the northern hemisphere being notably larger, as a result of the larger land masses \cite{IPCC13}

In this case, we need first to construct a different Green function for each latitude from the instantaneous $CO_2$ doubling experiments. Then, we perform the convolution of the Green functions with  the same ramp function and obtaining the prediction of the response to the $1\%$ per year increase in the $CO_2$ concentration for each latitude.

Figure \ref{Fig5} shows the result of our application of the response theory for predicting the response of the zonally averaged surface temperature to the considered forcing scenario: panel a) displays the projection performed using response theory, and panel b) shows the difference between the results obtained from the actual direct numerical simulations. We first  observe that the agreement is extremely good except  for the time window 20 - 60 y in the high latitudes of the Southern Hemisphere and 40 - 60 y in the high latitudes of the Northern Hemisphere, where the response theory underestimates the true amount of surface temperature increase.

Something interesting happens when looking at the latitudinal profile of the time horizons of escape from the statistics of the control run presented in Fig. \ref{Fig4}a given by $t_{min,2}^\tau$ and $t_{min,4}^\tau$, where $\tau$ is $70$ $y$. Interestingly, we find that, while the climate response is weaker in the tropics, one is able to detect climate change earlier than in the high latitude regions, the reason being that the interannual variability of the tropical temperature is much lower. Therefore, the signal-to-noise ratio is more advantageous. We need to note that our model does not feature processes responsible for important tropical variability like El-Ni\~no-Southern Oscillation (ENSO), so this result might be not so realistic, yet it seems to have some conceptual merit.

It is also rather attractive the fact that we are now able to relate to specific region the cold bias of the prediction for $\{T_S\}$ already seen in Fig. \ref{Fig3}. The reason for the presence of such discrepancies concentrated in the high latitude regions is relatively easy to ascertain. We can in fact attribute this exactly to the inability of a linear method like the one used here to represent accurately the strong nonlinear ice-albedo feedback, which dominates the climate response of the polar regions, and especially over the sea areas. 

\begin{figure}[ht]
a) \includegraphics[width=0.65\textwidth]{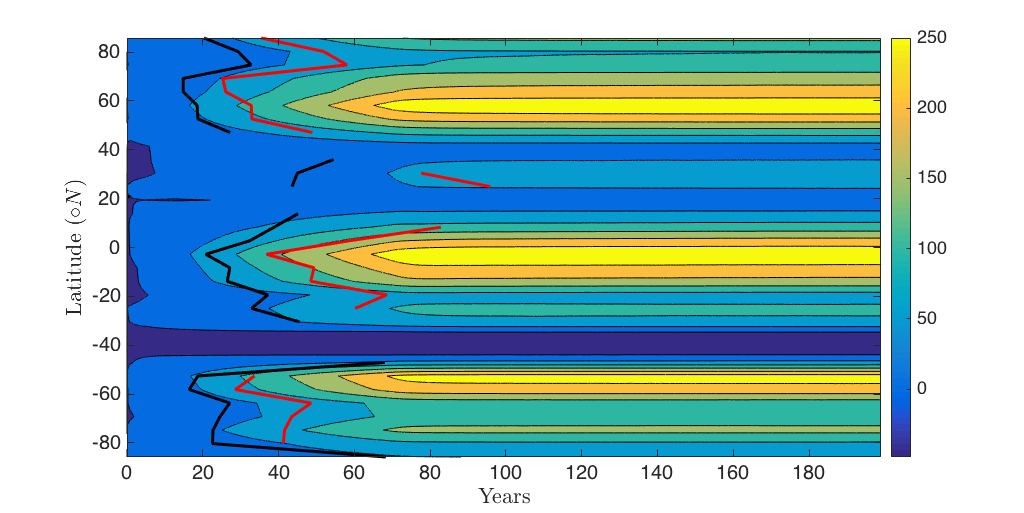}\\
b)  \includegraphics[width=0.65\textwidth]{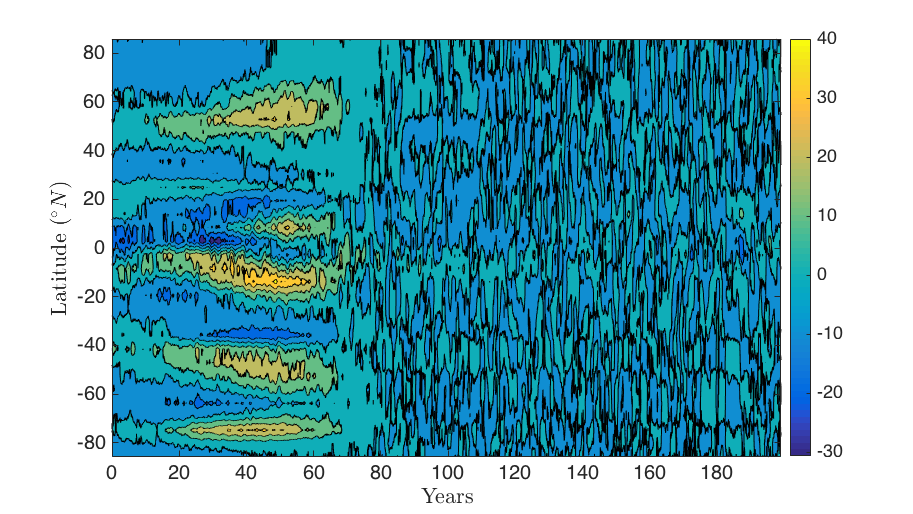}
\caption{Patterns of climate response - yearly total precipitation  $P_y$. a) Projection of the change of the zonal average of $P_y$ for different time horizons. The black and red lines indicates the escape times as presented in Fig \ref{Fig4} for $\tau=70$ b) Difference between the ensemble average of the direct numerical simulations and the predictions obtained using the response theory. }\label{Fig7}
\end{figure}

This can be made even more clear by looking at the performance of the response theory in predicting the 2D patterns of change of $T_S$. This is shown in Fig. \ref{Fig6}. where we see more clearly the geographical features of the changes in $T_S$ described above, and find confirmation that the sources of biases come exactly from the high-latitude sea-land margins, where sea ice is present. We also note that while for the time horizon of $20$ $y$ the bias between the simulations and the predictions of the response theory is comparable to the actual signal of response, the situation greatly improves for the time horizon of $60$ $y$. As we see here, there is good hope in being able to predict quite accurately the climate response also at local scale, with no coarse graining involved, at least in the case of the $T_S$ field.

As a final step, we wish to discuss climate projections for the quantity $[P_y]$. Figure \ref{Fig5b} shows that our model give a  picture of long-term climate change that is overall compatible with the findings of more complex and modern models: we foresee an increase of the precipitation in the tropical belt and in the regions of the storm tracks in the mid-latitude of the two hemisphere, whereas the change in small or negative in the remaining parts of the world. In other terms the regions that get a lot of rain are going to get even more, while drier regions do not benefit from the overall increase in the globally averaged precipitation. As we see, there is basically no correspondence between the patterns of change of $[P_y]$ and $[T_S]$ (despite the strong link between the response of the two globally averaged quantities) because precipitation patterns change as a result of a complex interplay of small and large scale processes, involving local thermodynamic exchanges, evaporation, as well as purely dynamical processes \cite{Allen2007}.

Panel a) of Figure \ref{Fig7} shows how response theory predicts the change in $[P_y]$ at different time horizons. We note that climate change is basically detectable only in the regions where strong increases of precipitation takes place, and the  horizon of escape from the control is much later in time compared to the case of $[T_S]$, the basic reason being that the variability of precipitation is much higher.

Panel b) of Fig. \ref{Fig7} shows the bias between the ensemble average of the numerical simulations and the prediction of the response theory. We notice that such biases are much stronger than in the case of $[T_S]$ (Fig. \ref{Fig5}b). In particular, we find rather interesting features in the time   horizon of $20-60$ $y$ for basically all latitudes. As we know from Fig. \ref{Fig3}b), the global average of such biases is rather small, but they are quite large locally. The projections performed using response theory underestimate the effects of some (relevant) nonlinear phenomena that impact the latitudinal distribution of precipitation, such as:
\begin{itemize}
\item Change in the relative size and symmetry of the ascending and descending branches of the Hadley cell, and in the position of its poleward extension: it is well known that strong warming can lead to a shift in the ITCZ, where where strong convective rain occurs, and to an extension of the dry areas of descending \cite{Chiang2012}. Correspondingly, the projections performed using the response theory are biased dry in the equatorial belt and biased wet in the subtropical band.
\item Impact on the water budget of the mid-latitudes of the increased water transport from the tropical regions taking place near the poleward extension of the Hadley cell, plus the change in the position of the stork track \cite{Yin2005}. As a result, the projection performed using the response theory is biased dry compared to the actual simulations in the mid-latitudes.
\end{itemize}

Comparing Figures \ref{Fig3}a), \ref{Fig3}b), \ref{Fig5}b), and \ref{Fig7}b) makes it clear that the performance of methods based on linear response theory depends strongly on the specific choice of the observable. In fact, when we choose an observable whose properties are determined by processes that are rather sensitive to our forcing, higher order corrections will be necessary to achieve a good precision.

\section{A Critical Summary of the Results}\label{conclusions}

This paper has been devoted to providing a statistical mechanical conceptual framework for studying the problem of climate change. We find it useful to construct the statistical properties of an unavoidably non-autonomous system like the climate using the idea of the pullback attractor and of the time-dependent measure supported on it. Response theory allows to practically compute such a time-dependent measure starting from the invariant measure of a suitably chosen reference autonomous dynamics.  

Using a the general circulation model PLASIM, we have developed response operators for predicting climate change resulting from an increase in the concentration of $CO_2$. The model features \textit{only} $O(10^5)$ degrees of freedom as opposed to $O(10^8)$ or more of state-of-the-art climate models \cite{IPCC13}, but, despite its simplicity, delivers a pretty good representation of Earth's climate and of its long-term response to $CO_2$ increase. PLASIM provides a low-resolution yet accurate representation of the dynamics of the atmosphere and of its coupling with the land surface, the ocean and the sea ice; it uses simplified but effective parametrizations for subscale processes including radiation, diffusion, dissipation, convection, clouds formation, evaporation, and precipitation in liquid and solid form. The main advantage of PLASIM is its flexibility and the relative low computer cost of launching a large ensemble of climate simulations. The main disadvantage is the lack of a representation of a dynamic ocean, which implies that we have a cut-off at the low frequencies, because we miss the multidecadal and centennial variability due to the ocean dynamics.  We have decided to consider classic IPCC scenarios of greenhouse forcing in order to make our results as relevant as possible in terms of practical implications.

The construction of the time dependent measure resulting from varying concentrations of $CO_2$ has been achieved by first performing a first set of simulations where $N=200$ ensemble members sampled from a long control run undergo an instantaneous doubling of the initial $CO_2$ concentration. Through simple numerical manipulations, we have been able to derive the linear Green function for any observable of interest, which makes it possible to perform predictions of climate change to an arbitrary pattern of change of the $CO_2$ concentration using simple convolutions, under the hypothesis that linearity is obeyed to a good degree of approximation.

We have studied the skill of the response theory in predicting the change in the globally averaged quantities as well as the spatial patterns of change to a forcing scenario of $1\%$ per year increase of $CO_2$ concentration up to doubling.  We have focused on observables describing the properties of two climatic quantities of great  geophysical interest, namely the surface temperature and the yearly total precipitation. The predictions of the response theory have been compared to the results of additional $N=200$ direct numerical simulations performed according this second scenario of forcing.

The performance of response theory in predicting the change in the globally averaged surface temperature and precipitation is rather good at all time horizons, with the predicted response falling within the ensemble variability of the direct simulations for all time horizons except for a minor discrepancy in the time window $40-60$ y. Additionally, our results confirm the presence of a strong linear link in the form of modified Clausius-Clapeyron relation between changes in such quantities, as already discussed in the literature. 

We have also studied how sensitive is the climate projection obtained using response theory to the size of the ensemble used for constructing the Green function. This is a matter of great practical relevance because it is extremely challenging to run a large number of ensemble members for specific scenarios using state-of-the-art climate models, given their extreme computational cost \cite{IPCC13}. We have then tested the skill of projections of globally averaged surface temperature and of globally averaged yearly total precipitation performed using Green functions constructed using $N=20$ ensemble members. We obtain that the quality of the projection is only moderately affected, and especially so in the case of the temperature observable.

By performing convolution of the Green function with various scenarios where the $CO_2$ increases at different rates, we are able to study the problem of climate change detection, associating to each rate of increase a time frame when climate change becomes statistically significant. Another new aspect of climate response we are able to investigate thanks to the methods developed here is the rigorous definition of transient climate sensitivity, which basically measures how different is the actual climate response with respect to the case of quasi-adiabatic forcings, and defines the thermal inertia of climate. 

We have shown that response theory allows to put in a broader and well defined context concepts like climate sensitivity:
\begin{itemize}
\item we understand that the equilibrium climate sensitivity relates to the zero-frequency response of the system to doubling of the $CO_2$ concentration: it is then clear that if we are not able to resolve the slowest time scales of the climate system, we will find  so-called state-dependency \cite{vdH2014,Koehler2015} when estimating \textit{equilibrium} climate sensitivity on slow (but not ultraslow) time scales, because we sample different regions of the climatic attractor;
\item we have that concepts like time-dependency \cite{Senior2000} of the equilibrium climate sensitivity are in fact related to the more concept of inertia of the climate response, which can be explored by generalizing the idea of transient climate sensitivity \cite{Allen2002} to all time scales of perturbations.
\end{itemize}

The analysis of the spatial pattern of climate change signal using response theory is entirely new and never attempted before. Clearly, when going from the global to local scale we have to expect lower signal-to-noise ration, as the variability is enhanced, and the possibility that linearity is a worse approximation as a result of powerful local nonlinear effects. Response theory provides an excellent tool also for predicting the change in the zonal mean of the surface temperature, except for an underestimation of the warming in the very high latitude regions in the time horizons of $40-60$ y. This is, in fact, the reason for the small bias found already when looking at the prediction of the globally averaged surface temperature. By looking at the 2D spatial patterns, we can associate such bias to a misrepresentation of the warming in the region where the presence of sea-ice is most sensitive to changing temperature patterns. The fact that linear response theory has problems in capturing the local features of a strongly nonlinear phenomenon like ice-albedo feedback makes perfect sense. In is remarkable that, instead, response theory is able to predict accurately the change in the surface temperature fields in most regions of the planet.

The prediction of the spatial patterns of change in the precipitation is much less successful, as a result of the complex nonlinear processes controlling the structure of the precipitative field. In particular, response theory is not able to deal effectively with describing the poleward shift of the storm tracks, in the widening of the Hadley cell, and in the change of the ICTZ. 

This paper provides a possibly convincing case for constructing climate change predictions in comprehensive climate models using concepts and methods of nonequilibrium statistical mechanics. The use of response theory potentially allows to reduce the need for running many different scenarios of climate forcings as in \cite{IPCC13}, and to derive, instead, general tools for computing climate change to \text{any} scenario of forcings from few, selected runs of a climate model. Additionally, it is possible to \textit{deconstruct} climate response to different sources of forcings  apart from increases in $CO_2$ concentration, \textit{e.g.} changes in $CH_4$ and aerosols concentration,  in land surface cover, in solar irradiance - and recombine it to construct very general climate change scenarios. While this operation is easier in a linear regime, it is potentially doable also in the nonlinear case, see \cite{LC12} for details. 

\section{Challenges and Future Perspectives}\label{future}

The limitations of this paper point at some potentially fascinating scientific challenges to be undertaken. Let's list some of them:
\begin{itemize}
\item A fundamental limitation of this study is the impossibility to resolve the centennial oceanic time scales. It is of crucial importance to test whether response theory is able to deal with prediction on a wider range of temporal scales, as required when ocean dynamics is included. We foresee future applications using a fully coupled yet efficient model like FAMOUS \cite{Jones2005}.
\item One should perform a systematic investigation of how appropriate linear approximation is in describing climate response to forcings, by computing estimates of the Green function using different level of $CO_2$ increases and testing them against a wide range of time modulating functions describing different scenarios of forcings.
\item It is necessary to study in greater detail what is the minimum size of the ensemble needed for achieving a good precision in the construction of the Green function; the requirement depends on the specific choice of the observable, including how it is constructed in terms of spatial and temporal averages of the actual climatic fields.
\item It is crucial  to look at the effect of considering multiple classes of forcings in the climate change scenarios and test how suitable combination of the individual Green functions associated to each separate forcing are able to predict climate response in general.
\end{itemize}

Different points of view on the problem of predicting climate response should as well be followed.  The \textit{ab initio} construction of the linear response operator has proved elusive because of the difficulties associated with dealing effectively with both the unstable and stable directions in the tangent space. It is promising to try to approach the problem by using the formalism of covariant Lyapunov vectors (CLVs) \cite{GPTCLP07,Wolfe2007,Kuptsov2012,Froyland2013} for having a convincing representation of the tangent space able to separate effectively and in an ordered manner the dynamics on the unstable and stable directions. CLVs have been recently shown to have great potential for studying instabilities and fluctuations in simple yet relevant geophysical systems \cite{Schubert2015}. By focusing on the contributions coming form the stable directions, one can also expect that such an approach might allow for estimating the - otherwise hard to control - error in the evaluation of the response operator introduced when applying the standard form of the fluctuation-dissipation theorem in the context of nonequilibrium systems possessing singular invariant measure. This would help understanding under which conditions climate-related applications of the fluctuation-dissipation theorem \cite{langen_estimating_2005,gritsun_climate_2007,majda07,gritsun2008b,cooper_climate_2011} have hope of being successful. 

{ One of the disadvantages of the CLVs is that constructing them is rather demanding in terms of computing power and requires a global (in time) analysis of the dynamics of the system, in order to ensure covariance, thus requiring relevant resources in terms of memory. Additionally, one expects that all the CLVs of index up to approximately the Kaplan-Yorke dimension of the attractor of the system are relevant for computing the response. Such a number, albeit typically much lower than the number of degrees of freedom, can still be extremely large for an intermediate complexity or, a fortiori, in a comprehensive climate model.  

At a more empirical level, a cheaper and effective alternative may be provided by the use of  Bred vectors (BVs). See comprehensive presentations in \cite{Deremble09,Norwood13}. BVs are finite-amplitude, finite time vectors constructed as the difference between  a background trajectory and a set of perturbed trajectories, where the perturbations (initially chosen at random having a small, yet finite norm) change following the nonlinear evolution of the trajectories, and are periodically rescaled to the prescribed initial norm. BVs provide a very efficient method for describing the main instabilities of the flow, taking into account nonlinear effects. In fact, what is extremely interesting about BVs is that a) their growth factors are strongly dependent on the region of the phase space where the system is; and b) the choice of the reference norm of the perturbation and of the time interval between two successive renormalization procedures (breeding period) effects strongly the properties of the dominant instabilities specifically active on the chosen time scales. Clearly, in the limit of infinitesimally small reference norm and infinitely long breeding time, all BVs converge to the first CLV. This is not the case when finite size effects become relevant. Instead, by considering not too small perturbations and a long enough time interval, the very fast instabilitities associated to the first CLVs are washed away by the loss of correlation due to nonlinear effects. In many applications of meteo-climatic relevance it has been shown that a relatively low number of BVs is extremely effective for reconstructiong tthe properties of the unstable space, and that BVs contain useful information on spatially localized features, so that it may be worth trying to construct an approximation to the Ruelle response operator using the BVs. One may note that considering different constructions of the BVs as discussed above might lead to the useful result of underlining different processes contributing to the response of the system. }

Another promising approach for studying climate response relies on reconstructing  the invariant measure of the unperturbed system using its unstable periodic orbits (UPOs) \cite{Cvitanovic1991}. Unstable periodic orbits has been shown to be a useful tool for studying persistent patterns and transitions in the context of simple atmospheric models \cite{Selten2004b,gritsun2008,gritsun2010}. Since resonances in the susceptibility describing the frequency-dependent response of specific observables can be associated to dominating UPOs (compare, \textit{e.g.}, \cite{L09} and \cite{bruno1994}), one can hope to be able to construct hierarchical approximations of the response operators by summing over larger and larger set of UPOs. 

Finally, it is worth mentioning that response theory can be approached in terms of studying the properties of the Perron-Frobenius transfer operator and of its generator  \cite{B00}  of the unperturbed and of perturbed system. In other terms, the focus is on studying directly how the invariant measure changes as a result of the applied perturbation and the challenge is in finding appropriate mathematical embedding  in terms of suitable functional spaces \cite{BL07,liverani2008,B08,B14b}. {See \cite{Ghil2015} for a proposal going in the direction of studying climate change by looking directly at measures rather than at observables, as instead done here.}

While the practical application of such an approach in a very high-dimensional problem like in the case of climate might in principle faces problems related to the curse of dimensionality, it has been advocated that it could provide an excellent framework for studying vicinity of the climate  to critical transitions \cite{chekroun2014}, \textit{i.e.} anticipating where there is no smoothness of the invariant measure with respect to perturbations. Such transitions are flagged by presence of rough dependence of the system properties on the perturbation due to presence of Ruelle-Pollicott resonances. This idea has been recently confirmed also analyzing long simulations performed with PLASIM and constructing a reduced space from two carefully selected observables \cite{Tantet2015b}. 
  
Recently, a comprehensive response theory for  Markov processes in a finite state space has been presented in the literature. Such a theory provides explicit matricial expressions of straightforward numerical implementation for constructing the linear and nonlinear response operators, including estimates of the radius of convergence \cite{L2015}. Using such results in a reduced state space might provide a novel and effective method for  approaching the problem of climate response.

\subsubsection*{Acknowledgements}
The authors wish to thank M. Chekroun, H. Dijkstra, M. Ghil, A. Gritsun, A. von der Heydt, and A. Tantet for sharing many ideas on climate sensitivity and climate response. Without such stimulations this work would have hardly been possible. The authors want to thank M. Ghil for providing a constructive and insightful review of this paper. VL wishes to thank D. Ruelle for sharing questions, answers, ideas, and doubts with such a special intellectual depth and generosity.

\end{document}